\documentclass[
  aps,
  prd,
  twocolumn,
  nofootinbib,
  superscriptaddress,
  floatfix,
  longbibliography
]{revtex4-2}

\usepackage{amsmath,amssymb,bm}
\usepackage{graphicx}
\usepackage{xcolor}
\usepackage{hyperref}

\graphicspath{{figures/}}
\hypersetup{
  colorlinks=true,
  citecolor=blue,
  linkcolor=blue,
  urlcolor=blue,
  pdftitle={Testing Dynamical Dark-Energy Model with Standard Sirens from the Einstein Telescope},
  pdfauthor={Yuanmin Mao and Wen Zhao},
  pdfkeywords={dark energy, gravitational waves, standard sirens, DESI DR2}
}

\newcommand{\dd}{\mathrm{d}}
\newcommand{\LCDM}{\ensuremath{\Lambda\mathrm{CDM}}}

\begin{document}

\title{Testing Dynamical Dark-Energy Model with Standard Sirens}

\author{Yuanmin Mao}
\email{maoyuanmin@mail.ustc.edu.cn}
\affiliation{Department of Astronomy, University of Science and Technology of
China, Hefei, Anhui 230026, People's Republic of China}

\author{Wen Zhao}
\email{wzhao7@ustc.edu.cn}
\affiliation{Department of Astronomy, University of Science and Technology of
China, Hefei, Anhui 230026, People's Republic of China}
\affiliation{School of Astronomy and Space Sciences, University of Science and
Technology of China, Hefei, Anhui 230026, People's Republic of China}
\affiliation{College of Physics, Guizhou University, Guiyang 550025, People's
Republic of China}

\begin{abstract}
Bright standard sirens from the Einstein Telescope (ET) provide an independent probe of cosmic expansion. In this article, we test whether future ET observations can constrain a dynamical dark-energy scenario motivated by DESI DR2. Using the best-fitting 29-bin dark-energy equation of state as the fiducial cosmology, we generate mock catalogs of 100, 500, 1000, and 3000 binary-neutron-star standard sirens with known host redshifts, and reconstruct the equation of state using both GW-only data and combined analyses with DESI DR2 BAO, BBN, and a compressed CMB constraint. The effective numbers of constrained modes are \(N_{\mathrm{eff}}\sim4-5\) (GW-only) and \(5-6\) (combined), with information concentrated in a few correlated modes. The GW3000 catalog yields a departure from \(\Lambda\)CDM exceeding \(5\sigma\), and combined analyses reach similar significance for \(\mathcal{O}(10^{3})\) events. Comparing flat \(\Lambda\)CDM with its Chevallier--Polarski--Linder extension using GW data alone gives significances of \(3.3\sigma\) and \(6.3\sigma\) for 1000 and 3000 events, rising to \(7.7\sigma\) and \(11.0\sigma\) with external data. These results indicate that a few thousand ET bright standard sirens could provide a sensitive test of departures from \(\Lambda\)CDM model.
\end{abstract}

\keywords{dark energy, gravitational-wave astronomy, standard sirens,
baryon acoustic oscillations}

\maketitle

\section{Introduction}
\label{sec:introduction}

The discovery of the accelerated expansion of the Universe established dark
energy as one of the central questions in modern cosmology
~\cite{Riess1998,Perlmutter1999}. The spatially flat $\Lambda$CDM model has
provided a remarkably successful description of a wide range of observations,
including measurements of the cosmic microwave background (CMB) and baryon
acoustic oscillations (BAO)~\cite{2020}. Nevertheless, the physical origin of
late-time cosmic acceleration remains unknown. A key observational question is
whether the expansion history is fully consistent with a cosmological constant
or whether it contains signatures of a dynamical dark-energy component.

Recent BAO measurements from the Dark Energy Spectroscopic Instrument (DESI)
Data Release 2 (DR2) have renewed interest in this question
~\cite{DESIDR2Cosmology,DESIDR2ExtendedDE}. By combining DESI DR2 BAO
measurements with Big Bang nucleosynthesis (BBN), compressed CMB acoustic-scale
constraints, and Type Ia supernovae (SNe Ia), Ref.~\cite{Gu_2025} reconstructed
the dark-energy equation of state in a nonparametric manner. In this analysis,
the dark-energy history was represented by 29 redshift bins over
$0<z\leq2.5$, with correlations between neighboring bins regularized through a
Horndeski-motivated prior~\cite{Raveri2017}. The resulting reconstruction
exhibits a redshift-dependent deviation from $\Lambda$CDM, providing a possible
indication of dynamical dark energy.

However, interpreting such a deviation requires independent confirmation from
complementary probes. In particular, the late-time distance information entering
current analyses relies significantly on SNe Ia. Although SNe Ia have played a
fundamental role in establishing cosmic acceleration, their cosmological
application involves empirical standardization relations based on light-curve
properties and spectral information~\cite{Phillips1993,Guy2007}. The absolute
magnitude calibration is usually marginalized as a nuisance parameter, while
residual uncertainties associated with photometric calibration, light-curve
modeling, and survey systematics may affect the inferred distance--redshift
relation~\cite{Scolnic2018,Brout2022}. Furthermore, standardized SNe Ia
luminosities exhibit correlations with host-galaxy properties, introducing
additional astrophysical uncertainties in precision cosmology
~\cite{Kelly2010}. Possible population evolution of SNe Ia with
redshift may also introduce systematic effects in the interpretation of their
cosmological distances~\cite{Childress2014}. 

Gravitational-wave (GW) standard sirens provide an independent route to
measure the expansion history with different systematic uncertainties
~\cite{Schutz1986,HolzHughes2005}. Unlike electromagnetic distance indicators,
the luminosity distance of a standard siren is directly encoded in the GW
waveform. When the source redshift is obtained from an electromagnetic
counterpart or host-galaxy identification, bright standard sirens provide a
direct measurement of the luminosity distance--redshift relation.
The observation of GW170817 demonstrated this approach and established the
foundation for GW-based cosmological measurements
~\cite{Abbott2017GW170817,Abbott2017StandardSiren}.

Future third-generation detectors, such as the Einstein Telescope (ET), are
expected to detect large populations of binary-neutron-star mergers and may
provide thousands of bright standard sirens extending to higher redshifts
~\cite{Zhao2011,ETDesign,Branchesi2023}. Such a large and independent
distance sample offers a unique opportunity to test whether the apparent
departure from $\Lambda$CDM inferred from current electromagnetic observations
can be reproduced by a fundamentally different cosmological probe.

In this work, we investigate the capability of ET bright standard sirens to
test the DESI DR2-inspired dynamical dark-energy history. We
adopt its best-fitting 29-bin $w(z)$ reconstruction as a fiducial scenario and
ask whether future GW observations can recover and distinguish this expansion
history from $\Lambda$CDM. We generate mock catalogs of
$N=100$, 500, 1000, and 3000 binary-neutron-star bright standard sirens with
known redshifts, including instrumental distance uncertainties and weak-lensing
scatter.

We perform two complementary analyses. First, combining ET catalogs with DESI
DR2 BAO, BBN, and compressed CMB constraints, we reconstruct the full
29-dimensional $w(z)$ history. Principal component analysis (PCA) is then used
to identify the redshift-dependent combinations constrained by the data. This
allows us to quantify the effective dimensionality of the measurable dark-energy
space and evaluate the covariance-weighted deviation from $\Lambda$CDM.
Second, we analyze the GW catalogs independently within a low-dimensional
framework by fitting flat $\Lambda$CDM and the Chevallier--Polarski--Linder
(CPL) parameterization~\cite{Chevallier2001,Linder2003}. Since the fiducial
history is intrinsically nonparametric, the inferred $(w_0,w_a)$ parameters
should be interpreted as an effective projection of the underlying evolution
rather than a complete description of the reconstructed $w(z)$ behavior. This
analysis provides a complementary assessment of whether ET standard sirens can
identify a departure from $\Lambda$CDM using a commonly adopted dark-energy
parameterization.

Together, these two approaches address different aspects of future GW
cosmology: the nonparametric analysis determines which redshift-dependent
features of the dark-energy history are observationally accessible, while the
CPL analysis quantifies the detectability of the deviation within a
low-dimensional phenomenological model. The central goal of this work is
therefore to assess whether the DESI DR2-motivated deviation from $\Lambda$CDM
can be independently tested by future standard-siren observations, how the
sensitivity scales with the size of the GW catalog, and which combinations of
$w(z)$ contain the dominant discriminating information.

The paper is organized as follows. Sec.~\ref{sec:de_framework} introduces
the cosmological framework, while Sec.~\ref{sec:gw_signal} describes the GW
signal and ET response. Sec.~\ref{sec:mock_data} presents the generation of
mock catalogs. Sec.~\ref{sec:external_data} describes the external cosmological
datasets. Sec.~\ref{sec:cosmological_inference} presents the inference
methodology, including nonparametric reconstruction of $w(z)$, principal
component analysis, and the parametric CPL analysis. We summarize our
conclusions in Sec.~\ref{sec:conclusions}.

\section{Dark-Energy Cosmological Framework}
\label{sec:de_framework}

We describe the background geometry by the
Friedmann--Lema\^{\i}tre--Robertson--Walker (FLRW) metric,
\begin{equation}
\dd s^2 = -\dd t^2 + a^2(t)\left[\frac{\dd r^2}{1-k r^2}+r^2\left(\dd\theta^2+\sin^2\theta\,\dd\phi^2\right)\right],
\end{equation}
where $t$ is cosmic time, $(r,\theta,\phi)$ are comoving spatial
coordinates, and $a(t)$ is the scale factor. The values $k=0$, 1, and
$-1$ describe flat, closed, and open geometries, respectively. We restrict
the analysis to $k=0$ in this article.

In general relativity, with minimally coupled radiation, baryons, cold
dark matter, and dark energy, the background expansion obeys
\begin{equation}
H^2 = \frac{8\pi G}{3}\rho_{\rm tot},
\label{eq:friedmann}
\end{equation}
\begin{equation}
\frac{\ddot{a}}{a}=-\frac{4 \pi G}{3}\left(\rho_{\mathrm{tot}}+3 p_{\mathrm{tot}}\right),
\end{equation}
where $H\equiv \dot a/a$, an overdot denotes differentiation with respect
to $t$, and $G$ is Newton's gravitational constant. The total energy
density and pressure are
\begin{equation}
\rho_{\rm tot}=\rho_r+\rho_b+\rho_c+\rho_{\rm de}, \qquad p_{\rm tot}=p_r+p_b+p_c+p_{\rm de},
\end{equation}
where the subscripts $r$, $b$, $c$, and ${\rm de}$ denote radiation,
baryons, cold dark matter, and dark energy. We also define
$\rho_m\equiv\rho_b+\rho_c$ and $p_m\equiv p_b+p_c$. Radiation is
negligible at the late times relevant here. Because the components interact
only gravitationally, each obeys the continuity equation
\begin{equation}
\dot\rho_i + 3H\left(1+w_i\right)\rho_i = 0.
\end{equation}
The equation-of-state parameter is $w_i\equiv p_i/\rho_i$. Matter has
$w=0$, whereas dark energy can have a redshift-dependent equation of
state, $w=w(z)$. For instance, in $\LCDM$, dark energy is a cosmological
constant with
\begin{equation}
w(z)=-1.
\end{equation}
In the CPL model, its equation of state is~\cite{Chevallier2001,Linder2003}
\begin{equation}
w(z)=w_0+w_a\frac{z}{1+z},
\end{equation}
where $w_0$ is the present-day value and $w_a$ controls the redshift
evolution. The choice $(w_0,w_a)=(-1,0)$ recovers $\LCDM$.

Using $a=(1+z)^{-1}$, the dark-energy density evolves as
\begin{equation}
\rho_{\rm de}(z)=\rho_{{\rm de},0}\exp\left[3\int_0^z \frac{1 + w(z')}{1+z'}\,\dd z'\right].
\label{eq:rho_de_evolution}
\end{equation}
Substituting the total energy density into Eq.~\eqref{eq:friedmann} and
neglecting late-time radiation gives
\begin{equation}
H(z)=H_0\left[\Omega_m(1+z)^3
+\Omega_{\rm de}\frac{\rho_{\rm de}(z)}{\rho_{{\rm de},0}}\right]^{1/2}.
\label{eq:Hz_from_components}
\end{equation}
Here $\Omega_m$ and $\Omega_{\rm de}$ are the present matter and
dark-energy density parameters, with $\Omega_{\rm de}=1-\Omega_m$ for a
flat universe. The luminosity distance is
\begin{equation}
d_L(z)=(1+z)d_C(z),
\label{eq:dl_from_H}
\end{equation}
where
\begin{equation}
d_C(z)=c\int_0^z \frac{\dd z'}{H(z')}.
\label{eq:dc_from_H}
\end{equation}

\section{Gravitational-Wave Signal and Einstein Telescope}
\label{sec:gw_signal}
Unlike standard candles, GWs from inspiraling compact binaries provide an
absolute measurement of luminosity distance~\cite{Schutz1986,HolzHughes2005}.
For a binary neutron star (BNS), the phase evolution constrains the chirp
mass, while the waveform amplitude encodes the luminosity
distance~\cite{Sathyaprakash_2009}. BNS mergers can therefore serve as
standard sirens. We use geometrized units, $G=c=1$, in the waveform and detector-response
calculations.

Ground-based interferometers measure the dimensionless strain induced by a
passing GW\@. In the transverse-traceless (TT) gauge, the response is
\begin{equation}
h(t)
=
F_{+}(\theta,\phi,\psi)\, h_{+}(t)
+
F_{\times}(\theta,\phi,\psi)\, h_{\times}(t),
\end{equation}
where the independent polarizations satisfy $h_{+}=h_{xx}=-h_{yy}$ and
$h_{\times}=h_{xy}=h_{yx}$. The functions $F_{+}$ and $F_{\times}$ are
the detector antenna patterns, $\psi$ is the polarization angle, and
$(\theta,\phi)$ gives the source position in the detector
frame~\cite{Cai_2017}.

ET consists of three $60^\circ$ interferometers arranged in an equilateral
triangle and rotated by $120^\circ$ with respect to one another. Their
antenna patterns are~\cite{Zhao2011}

\begin{equation}
\begin{aligned}
F_{+}^{(1)}(\theta, \phi, \psi)= & \frac{\sqrt{3}}{2}\left[\frac{1}{2}\left(1+\cos ^2(\theta)\right) \cos (2 \phi) \cos (2 \psi)\right. \\
& \left.-\cos (\theta) \sin (2 \phi) \sin (2 \psi)\right] \\
F_{\times}^{(1)}(\theta, \phi, \psi)= & \frac{\sqrt{3}}{2}\left[\frac{1}{2}\left(1+\cos ^2(\theta)\right) \cos (2 \phi) \sin (2 \psi)\right. \\
& \left.+\cos (\theta) \sin (2 \phi) \cos (2 \psi)\right] \\
F_{+, \times}^{(2)}(\theta, \phi, \psi)= & F_{+, \times}^{(1)}(\theta, \phi+2 \pi / 3, \psi) \\
F_{+, \times}^{(3)}(\theta, \phi, \psi)= & F_{+, \times}^{(1)}(\theta, \phi+4 \pi / 3, \psi).
\end{aligned}
\end{equation}

We model nonspinning, quasicircular BNS signals with the TaylorF2
frequency-domain waveform in the stationary-phase and restricted
post-Newtonian approximations~\cite{Blanchet2014}. The amplitude is kept at
leading Newtonian order, proportional to $f^{-7/6}$, and the phase is
computed through 3.5 post-Newtonian order. The Fourier transform
$\mathcal{H}(f)$ of $h(t)$ is
\begin{equation}
\begin{aligned}
\mathcal{H}(f)= & \mathcal{A} f^{-7 / 6} \exp \left\{i \left[2 \pi f t_0-\frac{\pi}{4}+2 \Psi\left(\frac{f}{2}\right)\right.\right. \\
& \left.\left.-\varphi_{(2,0)}\right]\right\},
\end{aligned}
\end{equation}
where the Fourier amplitude $\mathcal{A}$ is
\begin{equation}
\begin{aligned}
\mathcal{A}= & \frac{1}{d_L}\left[F_{+}^2\left(1+\cos ^2\iota\right)^2+4F_{\times}^2\cos ^2\iota\right]^{1/2} \\
& \times \sqrt{\frac{5 \pi}{96}}\,\pi^{-7 / 6}\mathcal{M}_{c,\mathrm{obs}}^{5 / 6}.
\end{aligned}
\end{equation}
Here $t_0$ is the coalescence time, and $\iota$ is the angle between the
orbital angular momentum and the line of sight. For component masses $m_1$
and $m_2$, the total mass is $M=m_1+m_2$, the symmetric mass ratio is
$\eta=m_1m_2/M^2$, and the source-frame chirp mass is
$\mathcal{M}_c=M\eta^{3/5}$. The detector-frame chirp mass is
$\mathcal{M}_{c,\mathrm{obs}}=(1+z)\mathcal{M}_c$. The functions $\Psi(f)$
and $\varphi_{(2,0)}$ describe the post-Newtonian phase evolution and the
polarization-dependent phase offset of the dominant harmonic,
respectively~\cite{Zhao2011,Sathyaprakash_2009}.

\section{Mock GW Catalogs}
\label{sec:mock_data}
We construct the fiducial expansion history from the nonparametric
reconstruction of $w(z)$ in Ref.~\cite{Gu_2025}. That reconstruction used
DESI DR2 BAO measurements, a BBN prior~\cite{Schoneberg2026}, and a
compressed Planck Public Release 4 (PR4) constraint on the CMB acoustic
scale~\cite{Rosenberg_2022}. It represented the equation of state by 29
piecewise-constant bins, uniformly spaced in scale factor over
$0<z\leq2.5$, and imposed a Horndeski-motivated correlation
prior~\cite{Raveri2017}. We use the released best-fitting amplitudes $w_i$
as the fiducial history:
\begin{equation}
w_{\rm fid}(z)=
\begin{cases}
w_i,
& z_{i,\min}<z\leq z_{i,\max},
\quad i=1,\ldots,29,\\
-1,
& z>2.5,
\end{cases},
\label{eq:w_fid_binned}
\end{equation}
where $z_{i,\min}$ and $z_{i,\max}$ are the boundaries of bin $i$. Setting
$w_{\rm fid}=-1$ at $z>2.5$ preserves the standard early-Universe
expansion history.

For the luminosity-distance calculation, this prescription amounts to
the discretization
\begin{equation}
w(z) \rightarrow \{w_1, w_2, \dots, w_{29}\}.
\end{equation}
This piecewise-constant form permits an analytic evaluation of the
dark-energy density in Eq.~\eqref{eq:rho_de_evolution} within each bin. For
$w(z)=w_i$ over $z\in[z_i,z_{i+1}]$,
\begin{equation}
\int_{z_i}^{z} \frac{1+w_i}{1+z'}\,\dd z'
=
(1+w_i)\ln\frac{1+z}{1+z_i},
\label{eq:piecewise_w_integral}
\end{equation}
which determines the binned evolution of $\rho_{\rm de}(z)$ and hence
$H(z)$.

\begin{figure}[t]
\centering
\includegraphics[width=\columnwidth]{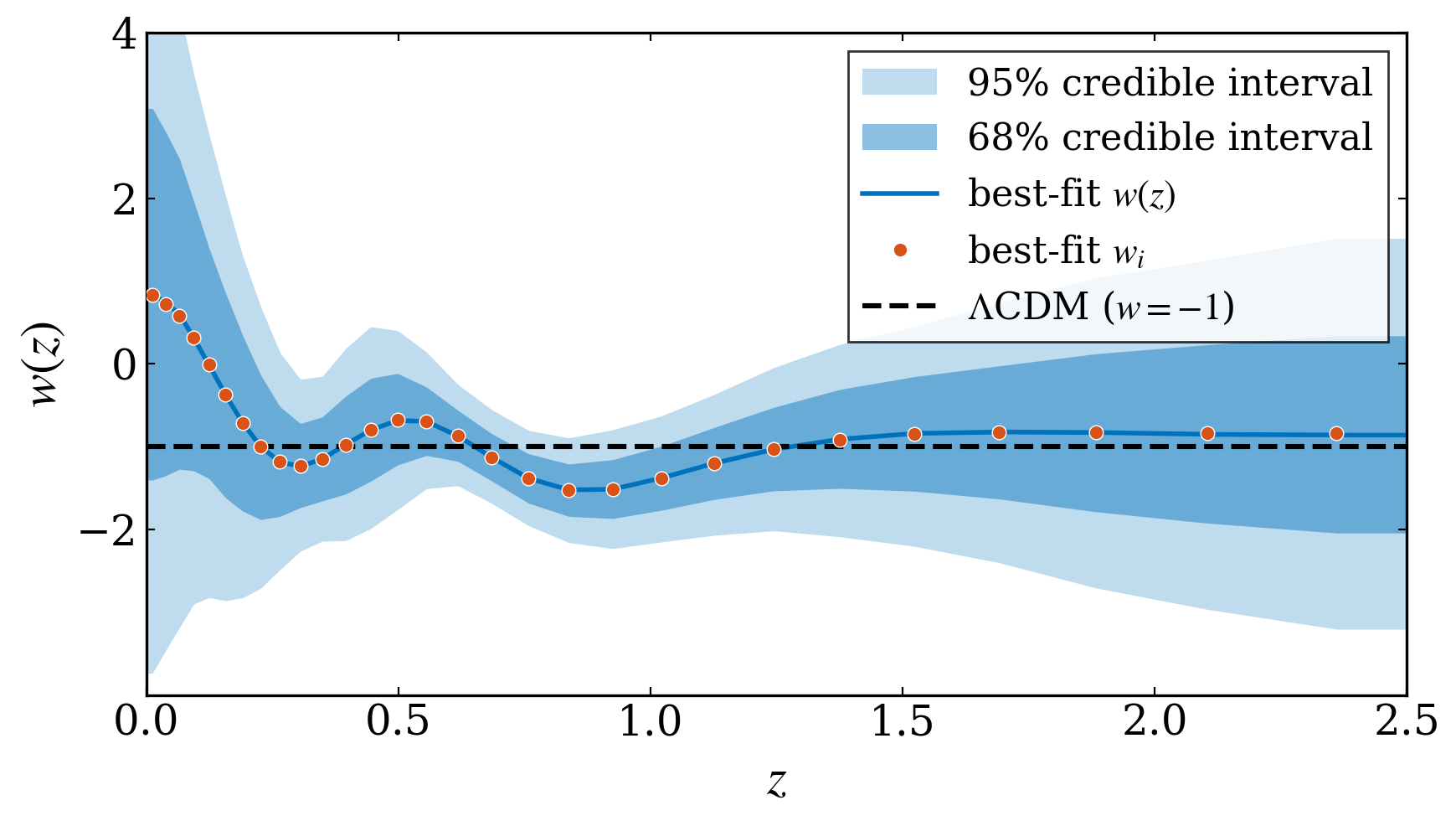}
\caption{Fiducial nonparametric dark-energy reconstruction adopted
from DESI DR2. Red points denote the best-fitting bin values $w_i$,
while the solid blue curve shows the corresponding reconstructed
$w(z)$. The shaded blue regions indicate the 68\% and 95\%
credible intervals, and the horizontal dashed black line
marks the $\LCDM$ prediction, $w=-1$.}
\label{fig:wz_posterior_and_cov}
\end{figure}

Figure~\ref{fig:wz_posterior_and_cov} shows the adopted history and its
uncertainty bands. Table~\ref{tab:wz_bins} lists the best-fitting bin values
used to calculate luminosity distances.

\begin{table}[t]
\caption{Best-fitting equation-of-state amplitudes for the fiducial DESI DR2
history~\cite{Gu_2025}. Within bin $i$, $w_{\rm fid}(z)=w_i$.}
\label{tab:wz_bins}
\centering
\small
\begin{ruledtabular}
\begin{tabular}{crcr}
$i$ & $w_i$ & $i$ & $w_i$ \\
\colrule
1  & \phantom{-}0.8320 & 16 & -0.8635 \\
2  & \phantom{-}0.7260 & 17 & -1.1272 \\
3  & \phantom{-}0.5847 & 18 & -1.3825 \\
4  & \phantom{-}0.3166 & 19 & -1.5248 \\
5  & -0.0088           & 20 & -1.5153 \\
6  & -0.3689           & 21 & -1.3809 \\
7  & -0.7129           & 22 & -1.1999 \\
8  & -0.9973           & 23 & -1.0302 \\
9  & -1.1787           & 24 & -0.9123 \\
10 & -1.2336           & 25 & -0.8460 \\
11 & -1.1538           & 26 & -0.8225 \\
12 & -0.9813           & 27 & -0.8234 \\
13 & -0.7915           & 28 & -0.8308 \\
14 & -0.6726           & 29 & -0.8371 \\
15 & -0.6911           &    &         \\
\end{tabular}
\end{ruledtabular}
\end{table}

To ensure consistency between the fiducial $w(z)$ history and the
background cosmology, we adopt the background parameters from the same
best-fit reconstruction: $H_0 = 57.365~\mathrm{km\,s^{-1}\,Mpc^{-1}}$ and
$\Omega_m = 0.423$ \cite{Gu_2025}.
We evaluate the dark-energy density with
Eq.~\eqref{eq:piecewise_w_integral} and obtain $H(z)$ from
Eqs.~\eqref{eq:rho_de_evolution} and \eqref{eq:Hz_from_components}. The
DESI-reconstructed expansion history then predicts the fiducial luminosity
distance $d_L^{\mathrm{fid}}(z)$ used to generate the mock catalogs.

Following Ref.~\cite{Zhao2011}, we model the redshift distribution of GW
sources as
\begin{equation}
P(z) \propto \frac{4\pi d_C^2(z) R(z)}{H(z)(1+z)},
\end{equation}
where the compact-binary merger rate $R(z)$ is~\cite{Cutler_2009}
\begin{equation}
R(z)=
\begin{cases}
1+2z, & z \le 1, \\
\frac{3}{4}(5-z), & 1 < z < 5, \\
0, & z \ge 5
\end{cases}.
\end{equation}

We model each luminosity-distance error using the standard ET forecasting
prescription. Because the GW amplitude scales as $d_L^{-1}$, the
Fisher-matrix instrumental error scales as $d_L/\rho$, where $\rho$ is the
ET network signal-to-noise ratio (SNR). We account conservatively for the
distance--inclination degeneracy by adopting \cite{Cai_2017}
\begin{equation}
\sigma_{d_L}^{\mathrm{inst}}
\simeq
\frac{2d_L}{\rho}.
\end{equation}

\begin{figure}[t]
\centering
\includegraphics[width=\columnwidth]{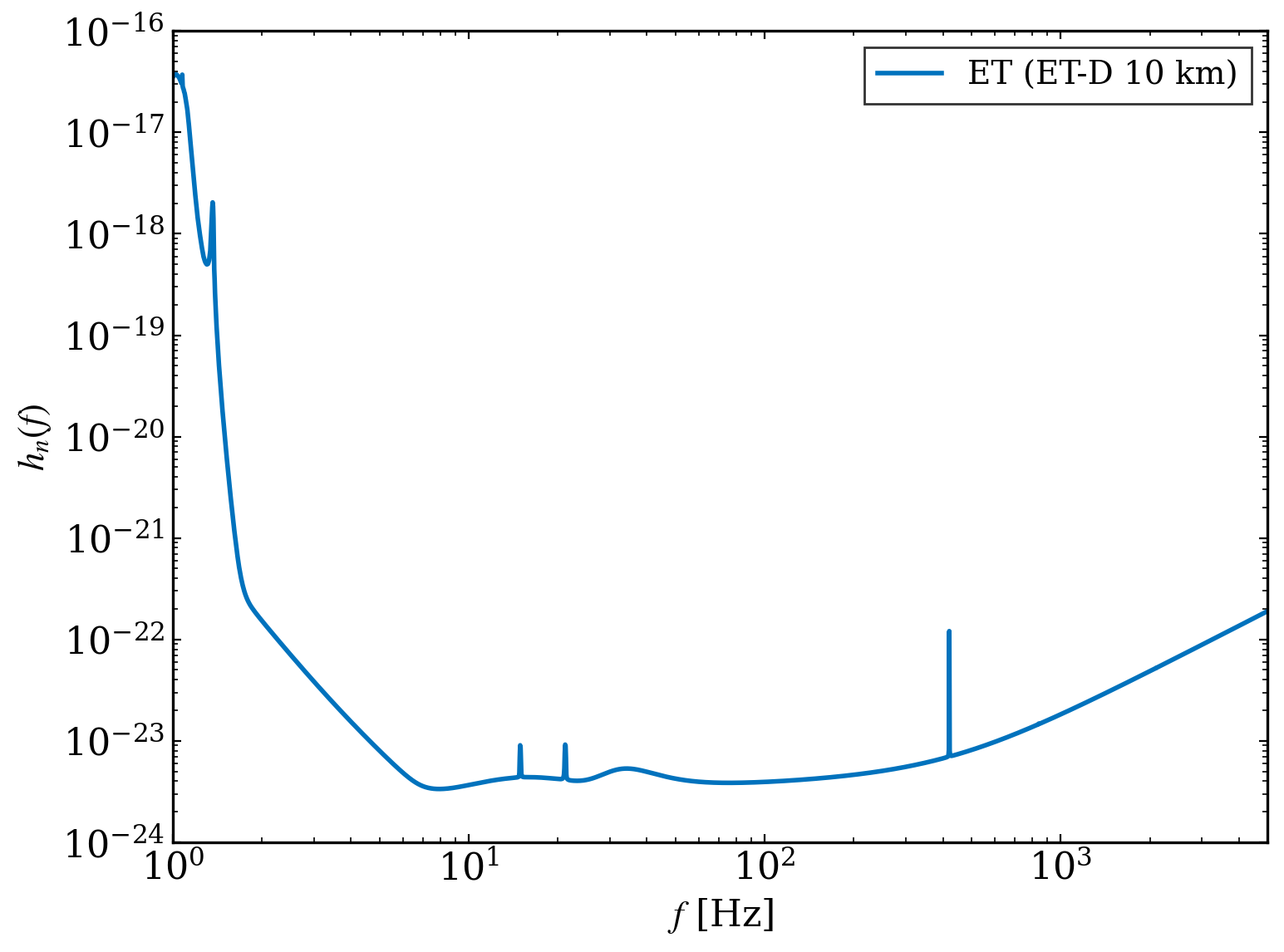}
\caption{Characteristic noise strain for the adopted ET-D sensitivity \cite{Hild_2011}.}
\label{fig:sensitivity_curves}
\end{figure}

The network SNR combines those of the three interferometers in quadrature,
\begin{equation}
\rho =
\left[
\sum_{j=1}^{3}
\left(\rho^{(j)}\right)^2
\right]^{1/2},
\end{equation}
where
$
\rho^{(j)}
=
\sqrt{
\left\langle
\mathcal{H}^{(j)},
\mathcal{H}^{(j)}
\right\rangle
},
$
denotes the matched-filter SNR of the $j$th
interferometer. The noise-weighted inner product is
defined as
\begin{equation}
\langle a,b\rangle
=
4
\int_{f_{\rm lower}}^{f_{\rm upper}}
\frac{
\tilde a(f)\tilde b^*(f)
+
\tilde a^*(f)\tilde b(f)
}{2S_h(f)}
\,\dd f,
\end{equation}
where $\tilde{a}(f)$ and $\tilde{b}(f)$ are Fourier transforms and $S_h(f)$
is the one-sided noise power spectral density. The characteristic noise
strain is $h_n(f)=\sqrt{fS_h(f)}$. We use the ET-D noise power spectral
density from Ref.~\cite{ETDesign}, shown in
Fig.~\ref{fig:sensitivity_curves}.

For each event, we draw the component masses independently from a Gaussian
distribution with mean $1.33\,M_\odot$ and standard deviation
$0.09\,M_\odot$, truncated to $[1.0,2.0]\,M_\odot$. This distribution is
consistent with the observed Galactic double-neutron-star
population~\cite{Ozel_2016,Farrow_2019}. We sample the sky
position $(\theta,\phi)$, polarization angle $\psi$, and inclination
$\iota$ isotropically. The SNR integral extends from
$f_{\rm lower}=1\,\mathrm{Hz}$ to $f_{\rm upper}=2f_{\rm LSO}$, where
$f_{\rm LSO}=1/(6^{3/2}2\pi M_{\rm obs})$ and
$M_{\rm obs}=(1+z)M$ is the detector-frame total mass. We retain events
with $\rho_{\rm net}>8$~\cite{Zhao2011,Cai_2017}. We
also include the weak-lensing error
$\sigma_{d_L}^{\rm lens}=0.05\,z\,d_L$~\cite{HolzLinder2005,Cai_2017}.
The total luminosity-distance uncertainty is
\begin{equation}
\sigma_{d_L} = \sqrt{(\sigma_{d_L}^{\mathrm{inst}})^2 + (\sigma_{d_L}^{\mathrm{lens}})^2}.
\end{equation}

\begin{figure*}[t]
\centering
\includegraphics[width=0.8\textwidth]{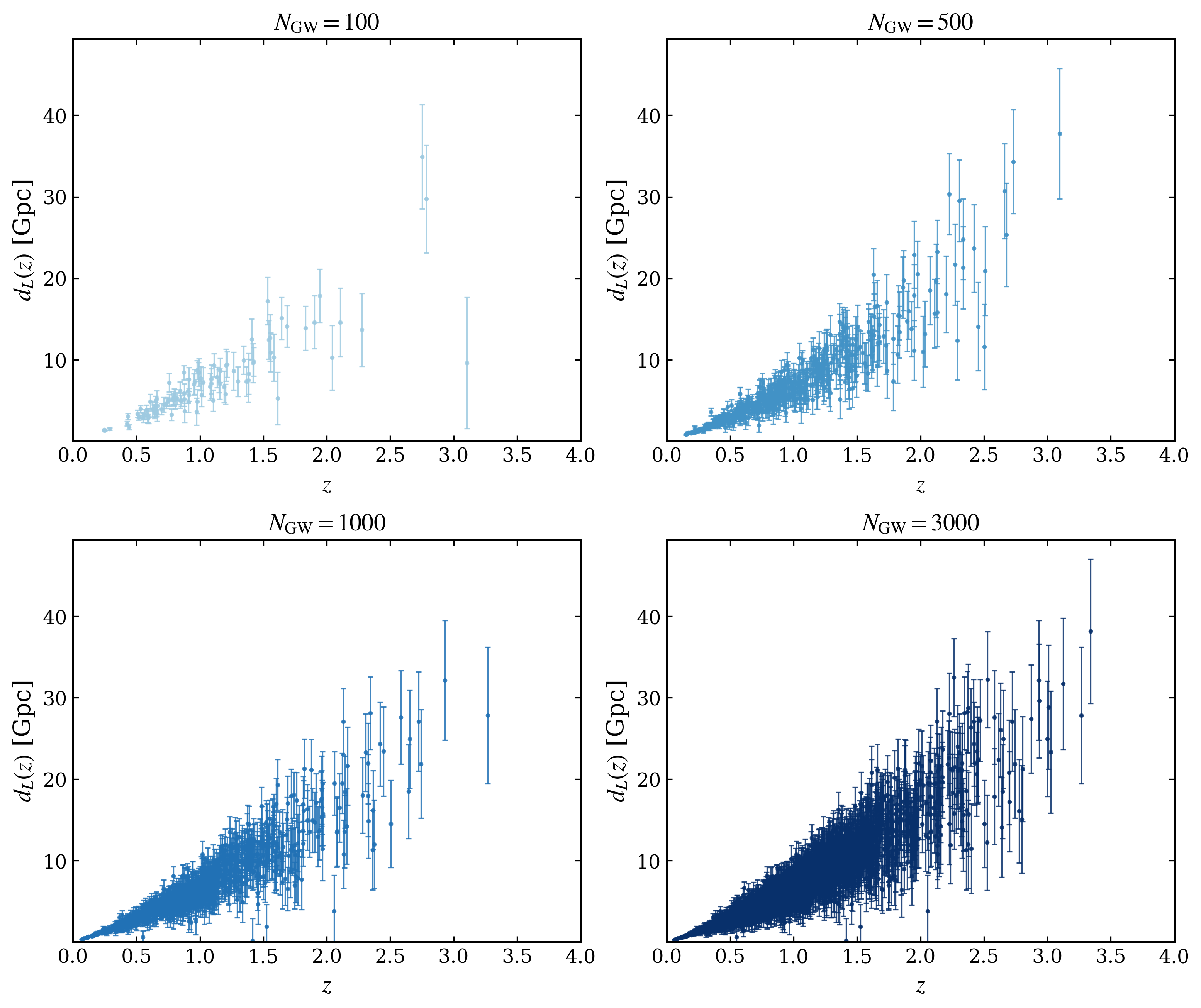}
\caption{
Mock ET standard-siren catalogs generated from the fiducial DESI DR2
reconstruction for $N=100$, 500, 1000, and 3000. Each point represents a
detected BNS with a redshift, luminosity distance, and total distance
uncertainty. The requirement $\rho_{\rm net}>8$ truncates the catalogs at
high redshift.
}
\label{fig:mock_GW_catalog_ET_by_4}
\end{figure*}

For each source, we draw a mock GW luminosity distance from
$d_{L,i}^{\mathrm{mock}}\sim
\mathcal{N}\!\left(d_{L,i}^{\mathrm{fid}},\sigma_{d_L,i}^{2}\right)$.
We generate catalogs of $N=100$, 500, 1000, and 3000
events with identified electromagnetic redshifts. Beyond the network-SNR
threshold, we impose no redshift-, inclination-, or sky-dependent
electromagnetic selection. We condition on a secure redshift and do not
predict an electromagnetic selection function or bright-siren
yield~\cite{ChenTalbotChase2024}. Thus, the catalog sizes count usable
bright standard sirens and do not correspond to fixed observing times.
Figure~\ref{fig:mock_GW_catalog_ET_by_4} shows the resulting
$\{z_i,d_{L,i}^{\mathrm{mock}},\sigma_{d_L,i}\}$.

\section{External Data}
\label{sec:external_data}
In this work, for some cases of analysis, we combine each mock GW catalog with
the following external data.

\subsection{DESI DR2 BAO}

We use 13 DESI DR2 BAO distance measurements spanning
$0.295\leq z_{\rm eff}\leq2.33$~\cite{DESIDR2Cosmology}.
They use four tracer types: the bright galaxy
sample (BGS)~\cite{Hahn_2023}, luminous red galaxies
(LRGs)~\cite{Zhou_2023}, emission-line galaxies
(ELGs)~\cite{Raichoor_2023}, and quasars
(QSO)~\cite{Chaussidon_2023}.
The data vector contains the isotropic BGS measurement $D_V/r_{\rm d}$ at
$z_{\rm eff}=0.295$ and six anisotropic pairs
$(D_M/r_{\rm d},D_H/r_{\rm d})$ from LRG1, LRG2,
LRG3$+$ELG1, ELG2, QSO, and Ly$\alpha$ (Table~IV of
Ref.~\cite{DESIDR2Cosmology}). The adopted measurements and their correlation
coefficients are summarized in
Table~\ref{tab:desi_dr2_bao}.

\begin{table*}[t]
\centering
\caption{
DESI DR2 BAO distance measurements adopted in this work.
The data consist of one isotropic measurement, $D_V/r_d$,
from BGS and six anisotropic measurement pairs,
$(D_M/r_d,D_H/r_d)$, from LRG, ELG, QSO, and
Ly$\alpha$ tracers~\cite{DESIDR2Cosmology}).
}
\label{tab:desi_dr2_bao}
\begin{ruledtabular}
\begin{tabular}{lccccc}
Tracer &
$z_{\rm eff}$ &
$D_V/r_d$ &
$D_M/r_d$ &
$D_H/r_d$ \\
\hline
BGS &
0.295 & $7.942\pm0.075$ & -- & --  \\
LRG1 &
0.510 & -- & $13.588\pm0.167$ &
$21.863\pm0.425$ \\
LRG2 &
0.706 & -- & $17.351\pm0.177$ &
$19.455\pm0.330$ \\
LRG3+ELG1 &
0.934 & -- & $21.576\pm0.152$ &
$17.641\pm0.193$ \\
ELG2 &
1.321 & -- & $27.601\pm0.318$ &
$14.176\pm0.221$  \\
QSO &
1.484 & -- & $30.512\pm0.760$ &
$12.817\pm0.516$  \\
Ly$\alpha$ &
2.330 & -- & $38.988\pm0.531$ &
$8.632\pm0.101$  \\
\end{tabular}
\end{ruledtabular}
\end{table*}

\subsection{BBN and CMB}

We impose a Gaussian BBN prior on the physical baryon density,
\begin{equation}
 \Omega_b h^2 = 0.02196 \pm 0.00063,
\end{equation}
following Ref.~\cite{Schoneberg2026}.

For the CMB, we use the compressed Planck PR4 acoustic-scale constraint
\begin{equation}
 100\theta_* = 1.04098 \pm 0.00042,
 \label{eq:theta_star_constraint}
\end{equation}
where
\begin{equation}
 \theta_* \equiv \frac{r_*}{D_M(z_*)}.
\end{equation}
Here $r_*$ is the comoving sound horizon at photon decoupling,
$z_*$ is the decoupling redshift, and $D_M(z_*)$ is the transverse
comoving distance to last scattering. Following Ref.~\cite{Gu_2025}, we
set $\sigma(100\theta_*)=0.00042$, a $75\%$ enlargement of the
\texttt{PR4\_12.6 TTTEEE} uncertainty in Table~5 of
Ref.~\cite{Rosenberg_2022}.

We assume standard $\LCDM$ physics before recombination and at the
baryon-drag epoch. We compute $r_{\rm d}$, $r_*$, and $z_*$
self-consistently with \texttt{CAMB}~\cite{Lewis_2000}.

\section{Cosmological Inference and Reconstruction Methods}
\label{sec:cosmological_inference}

In this section, we first define the common likelihood and sampling procedure, and then
describe the nonparametric reconstruction, PCA, effective dimensionality,
and complementary CPL model comparison.

\subsection{Cosmological inference}
\label{sec:likelihood_sampling}

We infer the cosmological parameters with the Metropolis--Hastings Markov
chain Monte Carlo (MCMC) sampler implemented in
\texttt{Cobaya}~\cite{Torrado2021}. For a simulated standard-siren catalog
$\{z_i,d_{L,i}^{\rm mock},\sigma_{d_L,i}\}_{i=1}^{N}$, the theoretical
luminosity distance is evaluated at the source redshifts,
$d_L^{\rm th}(z_i;\boldsymbol{\theta})$. In analysis, we assume the redshifts as known
and assume independent Gaussian distance errors. Therefore, the likelihood is
$\mathcal{L}\propto\exp(-\chi^2/2)$, where
\begin{equation}
 \chi^2_{\rm GW}(\boldsymbol{\theta}) =
 \sum_{i=1}^{N}
 \frac{\left[d_{L,i}^{\rm mock}
 -d_L^{\rm th}(z_i;\boldsymbol{\theta})\right]^2}
 {\sigma_{d_L,i}^{\,2}}.
 \label{eq:gw_chi2}
\end{equation}
When external probes are included, the corresponding DESI BAO, BBN, and CMB contributions 
are added to $\chi^2_{\rm GW}$ to construct the total data $\chi^2$, using the observables 
and uncertainties described in Sec.~\ref{sec:external_data}.
Table~\ref{tab:parameters_priors} lists the uniform parameter priors. A
parameter common to multiple models has the same prior in each model.

\begin{table}[t]
 \caption{Uniform parameter priors. The nonparametric analysis varies
 $H_0$, $\Omega_m$, $\Omega_b h^2$, and the 29 binned equation-of-state
 amplitudes $w_i$. The GW-only CPL analysis varies $H_0$, $\Omega_m$,
 $w_0$, and $w_a$. In flat $\LCDM$, $w_0=-1$ and $w_a=0$ are fixed.}
 \label{tab:parameters_priors}
 \centering
 \small
 \renewcommand{\arraystretch}{1.2}
 \setlength{\tabcolsep}{8pt}
 \begin{ruledtabular}
 \begin{tabular}{lc}
 Parameter & Prior \\
 \colrule
 $H_0\,[\mathrm{km\,s^{-1}\,Mpc^{-1}}]$ & $[20,100]$ \\
 $\Omega_m$ & $[0.01,0.99]$ \\
 $w_0$ & $[-3,3]$ \\
 $w_a$ & $[-30,3]$ \\
 \colrule
 $\Omega_b h^2$ & $[0.005,0.04]$ \\
 $w_i (i=1,\ldots,29)$ & $[-6,4]$ \\
 \end{tabular}
 \end{ruledtabular}
\end{table}

We require the chains to satisfy the Gelman--Rubin criterion
$R-1<0.01$~\cite{GelmanRubin} and verify an effective sample size greater
than $10^3$ for every sampled parameter. Marginalized posterior summaries
and credible regions are obtained with \texttt{GetDist}~\cite{Lewis_2019}.

\subsection{Nonparametric reconstruction of \texorpdfstring{$w(z)$}{w(z)}}
\label{sec:nonparametric_reconstruction}

Our primary analysis reconstructs $w(z)$ without imposing a prescribed
two-parameter redshift dependence. We adopt the same representation used
to generate the fiducial expansion history in Sec.~\ref{sec:mock_data}:
$w(z)$ is piecewise constant in 29 bins, uniformly spaced in scale factor
over $0<z\leq2.5$. The GW-only analysis samples
$\{H_0,\Omega_m,w_1,\ldots,w_{29}\}$, while the combined
DESI+BBN+CMB+GW analysis additionally includes
$\Omega_b h^2$ as a free parameter.

Because the distance observables constrain broad redshift integrals of
$w(z)$, the individual bin amplitudes are highly degenerate. We regularize
rapid bin-to-bin variations with the Horndeski-motivated correlation
covariance $\mathbf{C}_{\Pi}$~\cite{Raveri2017}. Its continuous form is
$C(a,a')=\sqrt{C(a)C(a')}R(a,a')$, with
$C(a)=0.05+0.8a^2$ and
$R(a,a')=\exp[-(|\ln a-\ln a'|/0.3)^{1.2}]$.
Following Refs.~\cite{Crittenden2012,Zhao_2017,Gu_2025}, we define a
five-bin moving-average matrix $\mathbf{S}$, with its weights renormalized
near the boundaries, and penalize departures from the locally smoothed
history through
\begin{equation}
 \chi^2_{\rm prior}
 =
 (\mathbf{w}-\mathbf{S}\mathbf{w})^{T}
 \mathbf{C}_{\Pi}^{-1}
 (\mathbf{w}-\mathbf{S}\mathbf{w}).
 \label{eq:chi2_prior}
\end{equation}
Equivalently, the prior precision in the $w$-bin space is
$\widetilde{\mathbf{C}}_{\Pi}^{-1}
=(\mathbf{I}-\mathbf{S})^T\mathbf{C}_{\Pi}^{-1}
(\mathbf{I}-\mathbf{S})$.

\begin{figure*}[t!]
\centering
\includegraphics[width=\textwidth]{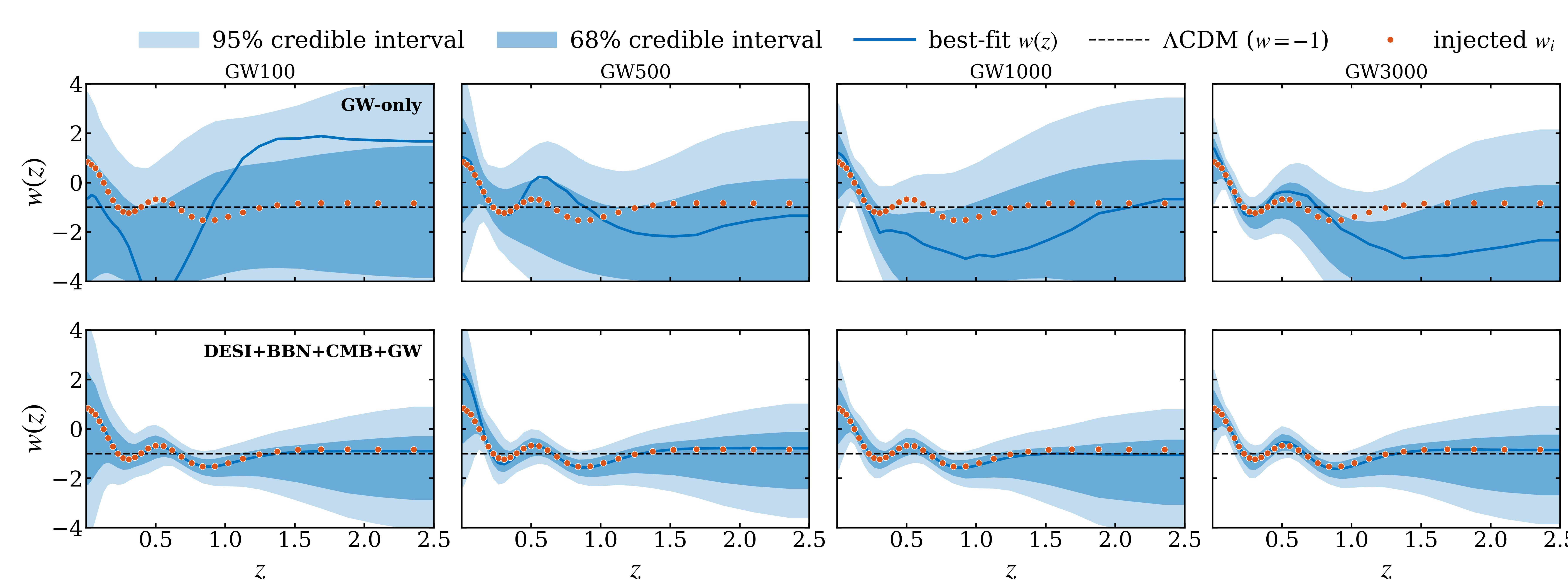}
\caption{Nonparametric reconstructions of the dark-energy equation of state obtained
from the GW-only and the combined DESI+BBN+CMB+GW analyses for mock
ET catalogs containing 100, 500, 1000, and 3000 standard sirens. The
dark- and light-blue bands show the 68\% and 95\% credible intervals,
respectively; the solid blue curves show the best-fitting reconstructions,
the orange points mark the injected bin values, and the dashed black lines
denote the flat-$\LCDM$ prediction $w=-1$.}
\label{fig:wz_posterior_cov_gw0100_0500_1000_3000_with_injected_w}
\end{figure*}

For the 32-dimensional parameter vector within the prior ranges of
Table~\ref{tab:parameters_priors}, the posterior is sampled using two types of catalogs with the correlation prior: (i) GW-only catalogs, and (ii) combined DESI+BBN+CMB+GW catalogs. 
\begin{equation}
 \chi^2
 =
 \chi_{\rm data}^2+A\chi^2_{\rm prior}.
 \label{eq:chi2_total_nonparam}
\end{equation}
The factor $A$ controls the strength of the correlation prior relative to
the data. Here, we adopt $A=1$ for the reconstructed $w(z)$ history. The results are shown in Fig.~\ref{fig:wz_posterior_cov_gw0100_0500_1000_3000_with_injected_w}. The GW-only catalogs provide limited constraints on the reconstructed $w(z)$ history due to the intrinsic distance-redshift degeneracy. Even for the largest catalog with 3000 standard sirens, the input $w(z)$ history can only be reliably recovered at low redshifts ($z\lesssim0.5$), while the reconstruction remains poorly constrained at higher redshifts. In contrast, the inclusion of DESI DR2 BAO, BBN, and the compressed CMB constraint significantly improves the reconstruction capability, leading to substantially tighter constraints over the low- and intermediate-redshift ranges. 
All combined DESI+BBN+CMB+GW analyses exhibit a consistent evolutionary pattern in $w(z)$, with $w>-1$ at $z\lesssim0.2$ and $w<-1$ around $z\simeq0.7–1.0$, closely following the injected dark-energy history.

\subsection{Principal components analysis}
\label{sec:pca}

The $w$ bins are strongly correlated, so the data constrain combinations of
bins rather than each bin independently. We use PCA to identify these modes~\cite{Huterer_2003},
\begin{equation}
 \mathbf{F}_w
 \equiv
 \mathbf{C}_w^{-1}
 =
 \mathbf{E}^{T}\boldsymbol{\Lambda}\mathbf{E},
 \quad
 \boldsymbol{\Lambda}
 =\operatorname{diag}(\lambda_1,\ldots,\lambda_{N_w}),
 \label{eq:pca_decomposition}
\end{equation}
where $N_w=29$, the rows of $\mathbf{E}$ are the orthonormal eigenmodes, and
larger $\lambda_i$ correspond to more tightly constrained modes. For each
catalog, $\mathbf{C}_w$ is the covariance matrix of the 29 $w$ bins for the
$A=1$ reconstruction, after marginalizing over $H_0$, $\Omega_m$, and
$\Omega_b h^2$.

We identify $\mathbf{w}_{\rm best}$ with the 29-component $w$-bin vector
from the converged MCMC sample having the minimum total effective $\chi^2$.
Relative to the $\LCDM$ reference history,
$\mathbf{w}_{\Lambda}=(-1,\ldots,-1)^T$, the reconstructed departure is
expanded as
\begin{equation}
 \mathbf{w}_{\rm best}-\mathbf{w}_{\Lambda}
 =
 \sum_{i=1}^{N_w}\gamma_i\mathbf{e}_i,
 \quad
 \gamma_i
 =
 \mathbf{e}_i^T
 \left(
 \mathbf{w}_{\rm best}-\mathbf{w}_{\Lambda}
 \right).
 \label{eq:pca_mode_amplitudes}
\end{equation}
The marginalized uncertainty of the $i$th mode is
$\sigma(\gamma_i)=\lambda_i^{-1/2}$, and its signed signal-to-noise ratio is
$s_i\equiv\gamma_i/\sigma(\gamma_i)=\gamma_i\sqrt{\lambda_i}$. The
cumulative reconstruction significance (that is, the signal-to-noise ratio, SNR)
over the first $m$ modes is
\begin{equation}
 \mathrm{SNR}(m)
 =
 \left[
 \sum_{i=1}^{m}s_i^2
 \right]^{1/2}.
 \label{eq:cumulative_snr}
\end{equation}
This cumulative SNR measures the overall significance of the departure from
the reference history contained in the first $m$ PCA modes.

We further evaluate the Bayesian evidence ratio between the
PCA-truncated nonparametric $w(z)$ model and $\Lambda$CDM. For a
model retaining the first $m$ principal components, the Bayes factor
is defined as
\begin{equation}
\ln B_{m,\Lambda}
=
\ln Z_{m}
-
\ln Z_{\Lambda{\rm CDM}},
\end{equation}
where $Z_m$ and $Z_{\Lambda{\rm CDM}}$ denote the Bayesian evidences
of the $m$-mode $w(z)$ model and the $\Lambda$CDM model, respectively.
By varying the number of retained modes, we assess how the Bayesian
preference for dynamical dark energy depends on the effective
dimensionality of the reconstruction. The Bayesian evidence is
computed using the nested-sampling algorithm implemented in
\texttt{dynesty}~\cite{Speagle2020}.

\begin{table*}[t]
\caption{
Comparison of covariance-weighted reconstruction significance and Bayesian
evidence ratios for two analyses. 
The quantities SNR and $\ln B_{m,\Lambda}$ are calculated using the first
$m$ principal components.
}
\label{tab:pca_snr_bayes_combined}

\centering

\textbf{GW-only}

\vspace{0.2cm}

\begin{ruledtabular}
\begin{tabular}{c|cccc|cccc}
& \multicolumn{4}{c|}{SNR}
& \multicolumn{4}{c}{$\ln B_{m,\Lambda}$}
\\
$m$
& GW100 & GW500 & GW1000 & GW3000
& GW100 & GW500 & GW1000 & GW3000
\\
\hline
1 & 0.058 & 1.364 & 3.194 & 6.844
& -0.003 & -0.258 & 3.859 & 21.386\\
2 & 0.103 & 1.399 & 3.209 & 7.073
& -0.007 & -0.346 & 3.550 & 21.445\\
3 & 0.108 & 1.442 & 3.279 & 7.109
& -0.009 & -0.398 & 3.701 & 20.965\\
4 & 0.172 & 1.444 & 3.296 & 7.129
& -0.020 & -0.435 & 3.712 & 20.801\\
5 & 0.173 & 1.444 & 3.298 & 7.161
& -0.022 & -0.438 & 3.624 & 20.947\\
6 & 0.191 & 1.493 & 3.298 & 7.167
& -0.026 & -0.450 & 3.621 & 20.894\\
7 & 0.193 & 1.510 & 3.299 & 7.170
& -0.028 & -0.488 & 3.563 & 20.874
\end{tabular}
\end{ruledtabular}

\vspace{0.5cm}

\textbf{DESI+BBN+CMB+GW}

\vspace{0.2cm}

\begin{ruledtabular}
\begin{tabular}{c|cccc|cccc}
& \multicolumn{4}{c|}{SNR}
& \multicolumn{4}{c}{$\ln B_{m,\Lambda}$}
\\
$m$
& GW100 & GW500 & GW1000 & GW3000
& GW100 & GW500 & GW1000 & GW3000
\\
\hline
1 & 1.062 & 2.856 & 5.854 & 9.402
& -1.507 & 1.669 & 14.267 & 40.880\\
2 & 1.965 & 3.823 & 7.577 & 11.072
& -0.684 & 4.211 & 24.582 & 56.315\\
3 & 2.053 & 3.824 & 7.696 & 11.169
& -0.651 & 3.958 & 24.830 & 56.780\\
4 & 2.079 & 3.961 & 7.868 & 11.365
& -0.752 & 4.279 & 25.975 & 58.716\\
5 & 2.097 & 4.297 & 7.880 & 11.373
& -0.779 & 4.997 & 25.870 & 58.596\\
6 & 2.155 & 4.351 & 7.913 & 11.412
& -1.283 & 5.048 & 25.851 & 58.351\\
7 & 2.173 & 4.360 & 7.937 & 11.412
& -1.351 & 4.372 & 25.296 & 58.049
\end{tabular}
\end{ruledtabular}
\end{table*}

Table~\ref{tab:pca_snr_bayes_combined} summarizes the covariance-weighted
reconstruction significance and Bayesian evidence obtained from the first
$m$ principal components for both the GW-only and the combined
DESI+BBN+CMB+GW analyses.

For both analyses, the deviation from $\LCDM$ is mainly captured by the
first few principal components, indicating that the dynamical features of
the injected $w(z)$ history are dominated by a small number of correlated
modes. For the GW-only catalogs, the constraining power increases rapidly
with the number of standard sirens. The GW3000 catalog reaches a
covariance-weighted significance of $\mathrm{SNR}=7.170$ and a Bayesian
evidence ratio of $\ln B_{7,\Lambda}=20.874$, while the GW1000 catalog
provides a moderate preference with $\mathrm{SNR}=3.299$ and
$\ln B_{7,\Lambda}=3.563$. This demonstrates that a sufficiently large
bright-siren catalog can independently constrain departures from
$\LCDM$, although the distance--redshift information alone provides
limited sensitivity for smaller catalogs.

The inclusion of DESI DR2 BAO, BBN, and CMB constraints substantially
enhances the sensitivity to the reconstructed dark-energy history. For the
GW500 catalog, retaining the first six principal components gives
$\mathrm{SNR}=4.351$ with $\ln B_{6,\Lambda}=5.048$. For catalogs reaching
$\mathcal{O}(10^3)$ events, the covariance-weighted significance exceeds
$5\sigma$ and the Bayesian evidence strongly favors the dynamical
dark-energy reconstruction when several degrees of freedom are retained.
In particular, for $m=4$, the Bayesian evidence reaches its maximum for
both GW1000 and GW3000, with $\ln B_{4,\Lambda}=25.975$ and $58.716$,
corresponding to $\mathrm{SNR}=7.868$ and $11.365$, respectively.

To isolate the modes constrained by the data, we perform an auxiliary MCMC
reconstruction with $A=0.01$ in Eq.~\eqref{eq:chi2_total_nonparam} and
denote the resulting data-posterior covariance of $\mathbf{w}_{\rm d}$ by
$\mathbf{C}_{\rm d}$. The data precision matrix is then estimated by
subtracting the residual prior contribution,
\begin{equation}
 \mathbf{F}_{\mathrm{data}}
 =\mathbf{C}_{\mathrm{d}}^{-1}
 -0.01\widetilde{\mathbf{C}}_{\Pi}^{-1}.
 \label{eq:data_fisher}
\end{equation}
We apply Eq.~\eqref{eq:pca_decomposition} separately to
$\mathbf{F}_{\mathrm{data}}$ and $\widetilde{\mathbf{C}}_{\Pi}^{-1}$ to obtain the
eigenvalues.

\begin{figure*}[t!]
\centering
\includegraphics[width=0.8\textwidth]{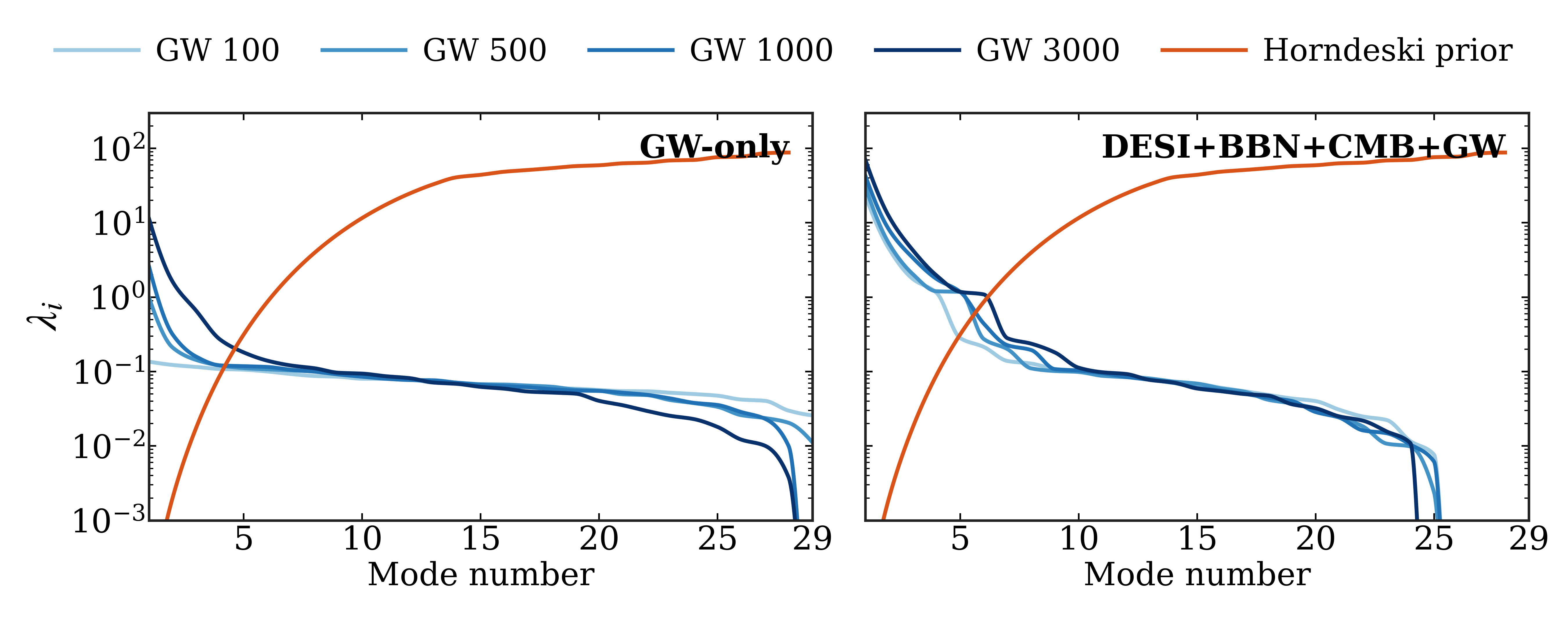}
\caption{
Principal-component precision spectra for the nonparametric
dark-energy reconstructions obtained from the GW-only and the combined
DESI+BBN+CMB+GW analyses. The blue curves show the data
eigenvalue spectra for different numbers of GW events, while the orange
curve shows the eigenvalue spectrum of the Horndeski correlation prior.
}
\label{fig:desi_dr2_bbn_cmb_absolute_gw_pca_spectrum}
\end{figure*}

Fig.~\ref{fig:desi_dr2_bbn_cmb_absolute_gw_pca_spectrum}
shows the principal-component precision spectra of the reconstructed
dark-energy equation of state for the GW-only and combined
DESI+BBN+CMB+GW analyses. The eigenvalues $\lambda_i$
quantify the constraining power of each principal component. 
For both analyses, the information is predominantly concentrated in the
first few principal components. Increasing the number of ET standard
sirens enhances the leading eigenvalues, indicating that larger GW
catalogs mainly improve the precision of the best-constrained correlated
combinations of the $w(z)$ history. However, the eigenvalues remain small
for most higher-order modes in the GW-only case, reflecting the limited
ability of luminosity-distance measurements alone to resolve the full
29-bin $w(z)$ parameter space. The inclusion of DESI DR2 BAO, BBN, and CMB
constraints significantly increases the leading eigenvalues by providing
additional information on the expansion history. The apparent truncation
near the 25th mode arises from several poorly constrained directions with
very small eigenvalues, corresponding to rapidly varying features between
neighboring bins that are suppressed by the Horndeski correlation prior
rather than independently measured by the data.

\begin{figure*}[t!]
\centering
\includegraphics[width=\textwidth]{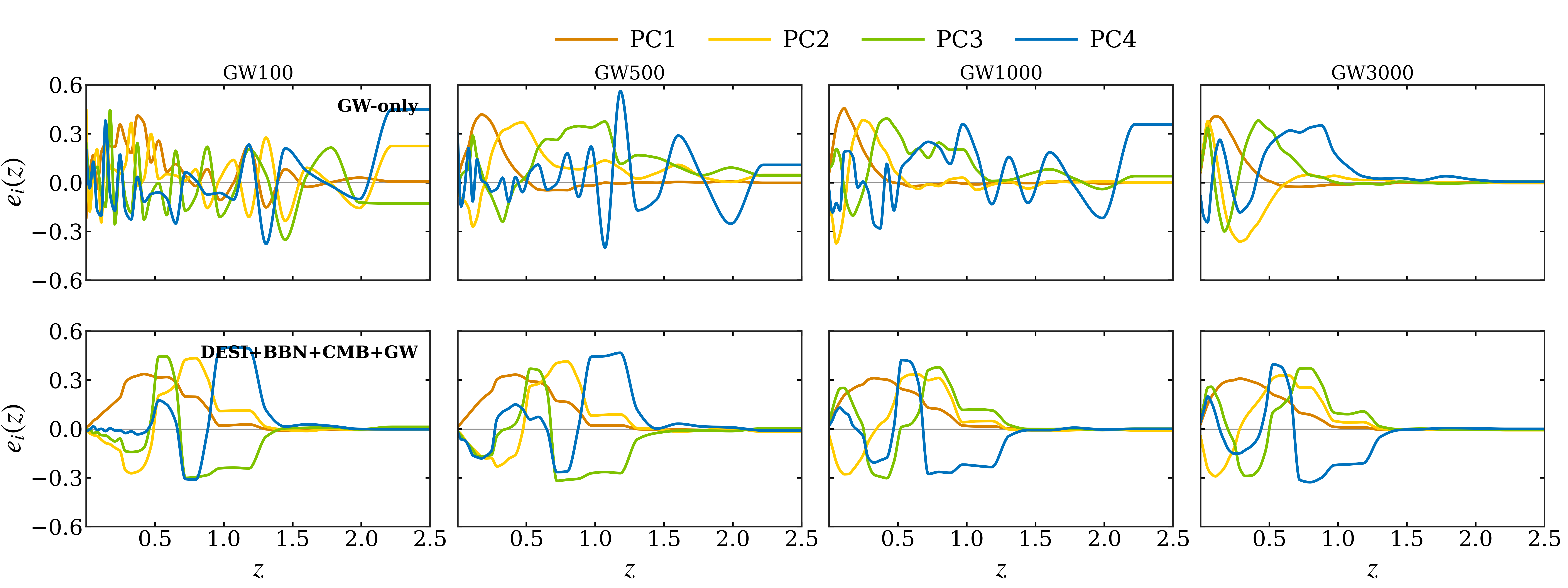}
\caption{The first four principal component eigenmodes for
the same combinations of the GW-only and the combined
DESI+BBN+CMB+GW analyses.
The four panels correspond to 100, 500, 1000, and 3000 GW events.}
\label{fig:wz_pca_modes_pc1-4_gw_only_vs_desi_bbn_cmb_gw0100_0500_1000_3000}
\end{figure*}

The corresponding eigenmodes in
Fig.~\ref{fig:wz_pca_modes_pc1-4_gw_only_vs_desi_bbn_cmb_gw0100_0500_1000_3000}
illustrate the redshift dependence of the correlated combinations constrained
by the data. In both the GW-only and combined analyses, the first principal
component (PC1) exhibits a broad and smooth structure, indicating that the
best-constrained combination mainly probes the overall variation of the
dark-energy history over a wide redshift range. Higher-order modes
(PC2--PC4) contain additional nodes and oscillatory features, corresponding
to more localized redshift-dependent variations that are progressively
less constrained.

For the GW-only analysis, the shapes of the leading eigenmodes become more
stable as the catalog size increases from 100 to 3000 events. In particular,
the small-scale oscillatory features present in the poorly constrained
higher-order modes of the GW100 case are gradually reduced, while the
dominant structures of PC1--PC4 are preserved. 
When DESI DR2 BAO, BBN, and CMB constraints are included, the eigenmodes are
more localized and better determined, especially for the higher-order
components. The external cosmological probes provide complementary
constraints on the expansion history and reduce the degeneracies among
different redshift bins. Nevertheless, the overall structure of the leading
principal components remains similar, suggesting that the additional
information primarily sharpens the measurement of the same dominant
$w(z)$ combinations.

Following Ref.~\cite{Raveri_2019}, we characterize the effective
number of independent $w(z)$ modes constrained beyond the correlation
prior by
\begin{equation}
 N_{\rm eff}
 =
 29-\operatorname{Tr}
 \left(
 \widetilde{\mathbf{C}}_{\Pi}^{-1}\mathbf{C}_{w}
 \right).
 \label{eq:neff}
\end{equation}

For the GW-only analysis, $N_{\rm eff}$ increases from
3.828 for GW100 to 5.191 for GW3000. With the inclusion of DESI DR2 BAO,
BBN, and the compressed CMB acoustic-scale constraint, the effective
dimensionality becomes $N_{\rm eff}=5.363$, $5.599$, $5.888$, and $6.274$
for GW100, GW500, GW1000, and GW3000, respectively. Although the external
probes substantially enhance the constraining power, the effective
dimensionality remains far below the original 29-dimensional parameter
space. Therefore, the reconstructed dark-energy history is effectively
described by only a small number of correlated modes, with increasing GW
statistics mainly improving their precision rather than opening many new
independent degrees of freedom.

\begin{figure*}[t!]
 \centering
 \includegraphics[width=0.78\textwidth]{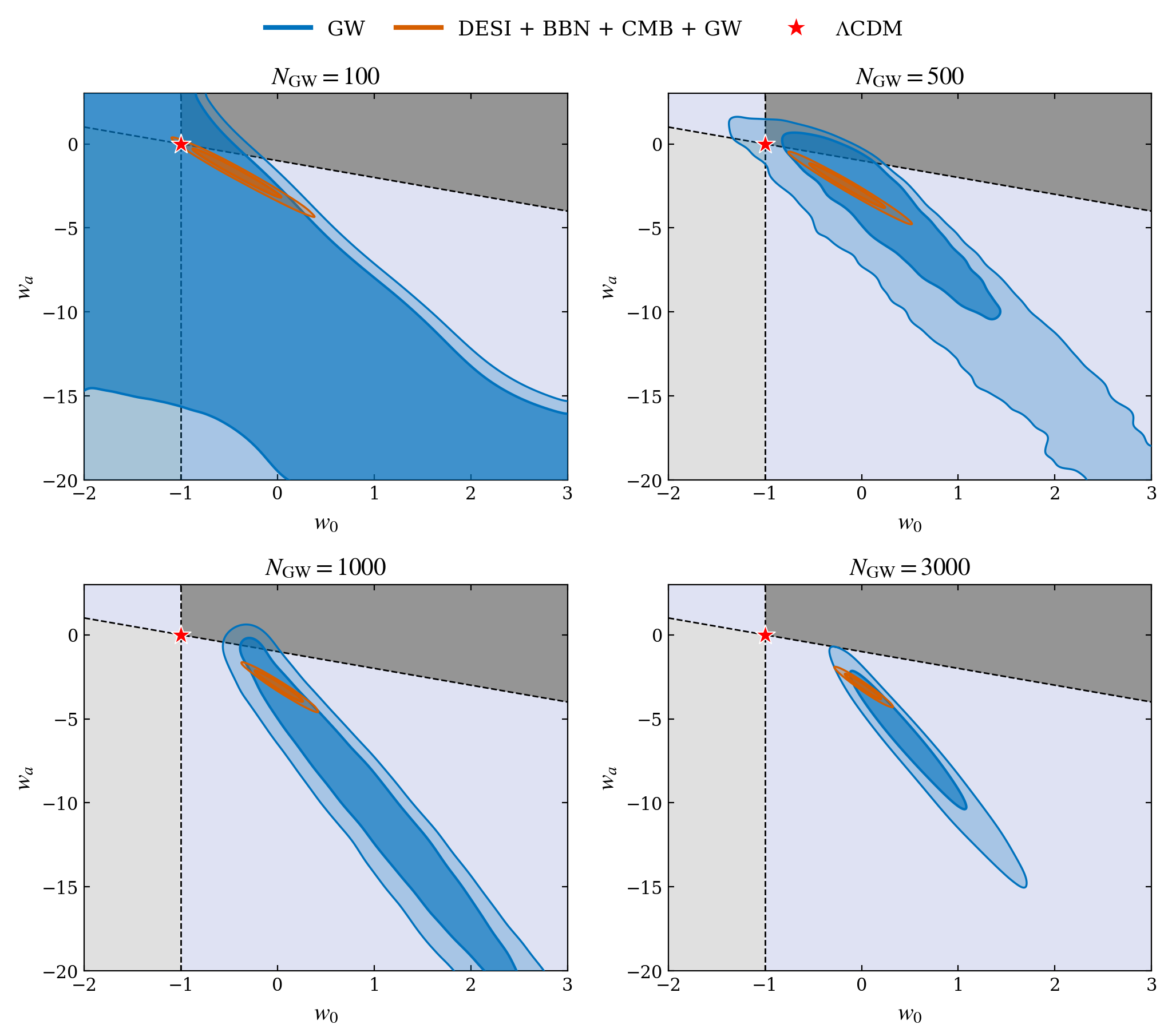}
 \caption{Marginalized 68\% and 95\% credible regions in the
$(w_0,w_a)$ plane from CPL fits to four representative ET
standard-siren catalogs with $N_{\rm GW}=100,500,1000,$ and $3000$.
The blue contours show constraints from the GW-only likelihood,
while the orange contours correspond to the combined DESI DR2 DESI+BBN+CMB+GW analysis. 
The red star marks the flat $\LCDM$ point,
$(w_0,w_a)=(-1,0)$.
The dashed lines indicate $w_0=-1$ and $w_0+w_a=-1$,
which separate the different CPL sectors.
The fitted CPL parameters provide an effective description of
the injected binned dynamical dark-energy history.}
\label{fig:gw_w0wa}
\end{figure*}

\subsection{Parametric CPL analysis}
\label{sec:cpl_inference}
As a complementary low-dimensional diagnostic, we fit each GW catalog under
spatially flat $\LCDM$ and its Chevallier--Polarski--Linder (CPL)
extension. Because the injected dark-energy history does not follow the CPL
form, the inferred $(w_0,w_a)$ values should be interpreted as an effective
projection of its luminosity-distance relation, rather than as a direct
recovery of the input $w(z)$. 
We consider two types of catalogs: (i) GW-only catalogs, where constraints are derived solely from standard siren measurements, and (ii) combined DESI+BBN+CMB+GW catalogs, where external cosmological probes are included.
The sampled parameter vectors of GW-only are
\begin{equation}
 \boldsymbol{\theta}_{\LCDM}=(H_0,\Omega_m),
 \quad
 \boldsymbol{\theta}_{\rm CPL}=(H_0,\Omega_m,w_0,w_a),
\end{equation}
while the combined
DESI+BBN+CMB+GW analysis additionally includes
$\Omega_b h^2$ as a free parameter.
The priors listed in Table~\ref{tab:parameters_priors}. These fits use
the GW likelihood in Eq.~\eqref{eq:gw_chi2}; one-dimensional constraints are
reported as marginalized posterior means and standard deviations.

We refine the maximum-likelihood estimates (MLEs) with
\texttt{iminuit}~\cite{Dembinski2020}, initializing the optimizer from the
highest-posterior samples of the independent chains. Among the converged solutions, we select the one with the minimum $\chi^2$. The
improvement over $\LCDM$ is quantified by
\begin{equation}
 \Delta\chi^2
 =\chi^2(\hat{\boldsymbol{\theta}}_{\LCDM})
 -\chi^2(\hat{\boldsymbol{\theta}}_{\rm CPL}),
 \label{eq:delta_chi2}
\end{equation}
where hats denote MLEs. Thus, $\Delta\chi^2>0$ indicates that CPL gives the
better maximum-likelihood fit.

Flat $\LCDM$ is nested within CPL at $(w_0,w_a)=(-1,0)$, and the two
models differ by two free parameters. Under the regularity conditions of
Wilks' theorem~\cite{Wilks1938}, $\Delta\chi^2$ follows a $\chi^2$
distribution with two degrees of freedom under the $\LCDM$ null
hypothesis. We quote a two-sided Gaussian-equivalent significance
$N_\sigma$ defined by
\begin{equation}
 \mathrm{CDF}_{\chi^2}\!\left(
 \Delta\chi^2\mid 2\,\mathrm{dof}
 \right)
 =
 \frac{1}{\sqrt{2\pi}}
 \int_{-N_\sigma}^{N_\sigma}
 e^{-t^2/2}\,\dd t.
 \label{eq:gaussian_equivalent_significance}
\end{equation}

\begin{table*}[t!]
\caption{
Comparison of the Wilks-theorem model-comparison statistics for the
GW-only and DESI+BBN+CMB+GW analyses.
The significance $N_\sigma$ is obtained from $\Delta\chi^2$
assuming two additional degrees of freedom ($\Delta k=2$).
For the GW-only case, the CPL maximum-likelihood solutions for
100 and 500 events reach the prior boundaries, and therefore the
corresponding significances are not reported.
}
\label{tab:wilks_comparison}
\centering
\small
\begin{ruledtabular}
\begin{tabular}{c|cc|cc}
& \multicolumn{2}{c|}{GW-only}
& \multicolumn{2}{c}{DESI+BBN+CMB+GW}
\\
$N_{\rm GW}$
& $\Delta\chi^2$
& $N_\sigma$
& $\Delta\chi^2$
& $N_\sigma$
\\
\hline
100
& 0.629
& --
& 4.348
& 1.582
\\
500
& 7.950
& --
& 16.694
& 3.676
\\
1000
& 13.944
& 3.309
& 64.010
& 7.710
\\
3000
& 44.086
& 6.317
& 127.004
& 11.033
\end{tabular}
\end{ruledtabular}
\end{table*}

Figure~\ref{fig:gw_w0wa} shows the marginalized CPL constraints in the
$(w_0,w_a)$ plane for the four representative ET standard-siren catalogs.
The blue contours correspond to the GW-only likelihood, while the orange
contours include the combined DESI+BBN+CMB+GW constraints.

For the GW-only analysis, the constraints exhibit a strong anticorrelation
between $w_0$ and $w_a$, reflecting the degeneracy of the CPL
parameterization. For 100 events, the broad posterior contains the
$\LCDM$ point $(w_0,w_a)=(-1,0)$. With 500 standard sirens, the allowed
region becomes narrower, although $\LCDM$ remains compatible with the
95\% credible region. For the 1000- and 3000-event catalogs, the posterior
shifts away from $\LCDM$ and is concentrated in the Quintom-B sector,
defined by $w_0>-1$ and $w_0+w_a<-1$~\cite{CaiQuintom2010}. These CPL
parameters should be interpreted as an effective low-dimensional
description of the injected binned equation-of-state history rather than
the original $w(z)$ bins themselves.
The inclusion of DESI+BBN+CMB data further tightens the CPL constraints
and enhances the displacement from $\LCDM$. This improvement reflects the
complementary information provided by early- and late-universe probes
when combined with standard sirens.

Table~\ref{tab:wilks_comparison} quantifies this displacement using the
Wilks likelihood-ratio statistic. For the GW-only case, the 100- and
500-event CPL maximum-likelihood solutions reach the prior boundaries,
and therefore the corresponding Wilks significances are not reported.
For the 1000- and 3000-event catalogs, the nominal significances are
$3.309\sigma$ and $6.317\sigma$, respectively. When DESI+BBN+CMB
constraints are included, the corresponding values increase to
$7.710\sigma$ and $11.033\sigma$. These results demonstrate that, for
the injected dynamical-dark-energy history and the adopted mock
realizations, future ET standard sirens combined with external
cosmological probes can substantially improve the discrimination between
flat $\LCDM$ and its effective CPL extension.

\section{Conclusions}
\label{sec:conclusions}

We have investigated the capability of Einstein Telescope (ET) bright
standard sirens to test a dynamical-dark-energy expansion history motivated
by DESI DR2. We adopted the best-fitting 29-bin reconstruction of $w(z)$
from Ref.~\cite{Gu_2025} as a fiducial cosmology and generated mock
binary-neutron-star bright standard-siren catalogs with identified host
redshifts. The analysis was performed in two complementary frameworks:
a nonparametric reconstruction of the dark-energy history using the
29-bin $w(z)$ parameterization, and a parametric comparison between flat
$\LCDM$ and the CPL extension using the GW likelihood.

The nonparametric analysis demonstrates that ET standard sirens constrain
a small number of correlated redshift modes rather than individual
$w(z)$ bins. The PCA analysis shows that the leading principal components
contain most of the information in both the GW-only and the combined DESI+BBN+CMB+GW analyses. Increasing the number of standard sirens mainly
improves the precision of these already constrained modes, while the
overall structures of the leading eigenmodes remain stable. For the
GW-only analysis, the effective dimensionality increases from
$N_{\rm eff}=3.828$ for GW100 to $5.191$ for GW3000. When DESI DR2 BAO,
BBN, and the compressed CMB acoustic-scale constraint are included, the
corresponding values become
$N_{\rm eff}=5.363$, $5.599$, $5.888$, and $6.274$ for GW100, GW500,
GW1000, and GW3000, respectively. Therefore, even with thousands of
standard sirens, the reconstructed 29-dimensional dark-energy history is
effectively described by only a small number of correlated modes.

The covariance-weighted reconstruction significance and Bayesian evidence
further quantify the departure from $\LCDM$. For the GW-only catalogs, the constraining power increases rapidly with
the number of standard sirens. For the GW3000 catalog, the cumulative
reconstruction significance exceeds $5\sigma$ and the Bayesian evidence
satisfies $\ln B_{m,\Lambda}>10$ for all retained modes from $m=1$ to $7$,
indicating a strong preference for the dynamical-dark-energy reconstruction
over $\LCDM$. In contrast, smaller catalogs provide weaker statistical
evidence due to their limited distance--redshift information. The inclusion of DESI DR2 BAO, BBN,
and CMB constraints significantly enhances the sensitivity. For example,
the GW500 catalog reaches
$\mathrm{SNR}=4.351$ and $\ln B_{6,\Lambda}=5.048$ when the first six
principal components are retained, and the GW1000 and GW3000 catalogs
achieve $\mathrm{SNR}=7.868$ and $11.365$, respectively, with the maximum
Bayesian evidence obtained at four retained modes. These results indicate
that the deviation from $\LCDM$ induced by the fiducial dynamical-dark-energy
history is primarily encoded in a few broad correlated modes.

The GW-only CPL analysis provides a complementary low-dimensional
description of the same expansion history. Since the injected $w(z)$
evolution is not restricted to the CPL form, the recovered
$(w_0,w_a)$ parameters should be interpreted as an effective projection of
the luminosity-distance relation rather than a direct reconstruction of
the original 29-bin history. For 100 and 500 standard sirens, the CPL
maximum-likelihood solutions reach the prior boundaries, preventing a
reliable Wilks-theorem significance conversion. For 1000 and 3000 events,
the nominal Gaussian-equivalent significances are
$3.31\sigma$ and $6.32\sigma$, respectively, under the two-degree-of-freedom
Wilks approximation.

When DESI BAO, BBN, and CMB acoustic-scale constraints are combined with
the GW catalogs, the CPL constraints become substantially tighter, with
nominal significances increasing to $7.71\sigma$ and $11.03\sigma$ for
the 1000- and 3000-event catalogs. The corresponding CPL parameters remain
in the Quintom-B sector, characterized by $w_0>-1$ and
$w_0+w_a<-1$, showing that this effective two-parameter description
captures the crossing-like behavior of the injected dynamical-dark-energy
history.

Under the assumptions of the adopted DESI DR2-motivated fiducial history,
the idealized bright-standard-siren catalogs, and the forecast ET distance
precision, these results demonstrate that future ET observations can
provide an independent probe of departures from the flat $\LCDM$ expansion
history. Combining standard sirens with complementary cosmological probes
will be essential for maximizing their capability to test dynamical dark
energy.

\section*{Acknowledgements}

This work is supported by the National Natural Science Foundation of China (grant No. 12325301), the Fundamental and Interdisciplinary Disciplines Breakthrough Plan of Ministry of Education of China and the National Key R\&D Program of China
(grant No. 2022YFC2204602).

%


\begin{thebibliography}{46}%
\makeatletter
\providecommand \@ifxundefined [1]{%
 \@ifx{#1\undefined}
}%
\providecommand \@ifnum [1]{%
 \ifnum #1\expandafter \@firstoftwo
 \else \expandafter \@secondoftwo
 \fi
}%
\providecommand \@ifx [1]{%
 \ifx #1\expandafter \@firstoftwo
 \else \expandafter \@secondoftwo
 \fi
}%
\providecommand \natexlab [1]{#1}%
\providecommand \enquote  [1]{``#1''}%
\providecommand \bibnamefont  [1]{#1}%
\providecommand \bibfnamefont [1]{#1}%
\providecommand \citenamefont [1]{#1}%
\providecommand \href@noop [0]{\@secondoftwo}%
\providecommand \href [0]{\begingroup \@sanitize@url \@href}%
\providecommand \@href[1]{\@@startlink{#1}\@@href}%
\providecommand \@@href[1]{\endgroup#1\@@endlink}%
\providecommand \@sanitize@url [0]{\catcode `\\12\catcode `\$12\catcode
  `\&12\catcode `\#12\catcode `\^12\catcode `\_12\catcode `\%12\relax}%
\providecommand \@@startlink[1]{}%
\providecommand \@@endlink[0]{}%
\providecommand \url  [0]{\begingroup\@sanitize@url \@url }%
\providecommand \@url [1]{\endgroup\@href {#1}{\urlprefix }}%
\providecommand \urlprefix  [0]{URL }%
\providecommand \Eprint [0]{\href }%
\providecommand \doibase [0]{https://doi.org/}%
\providecommand \selectlanguage [0]{\@gobble}%
\providecommand \bibinfo  [0]{\@secondoftwo}%
\providecommand \bibfield  [0]{\@secondoftwo}%
\providecommand \translation [1]{[#1]}%
\providecommand \BibitemOpen [0]{}%
\providecommand \bibitemStop [0]{}%
\providecommand \bibitemNoStop [0]{.\EOS\space}%
\providecommand \EOS [0]{\spacefactor3000\relax}%
\providecommand \BibitemShut  [1]{\csname bibitem#1\endcsname}%
\let\auto@bib@innerbib\@empty
\bibitem [{\citenamefont {Riess}\ \emph {et~al.}(1998)\citenamefont {Riess}
  \emph {et~al.}}]{Riess1998}%
  \BibitemOpen
  \bibfield  {author} {\bibinfo {author} {\bibfnamefont {A.~G.}\ \bibnamefont
  {Riess}} \emph {et~al.},\ }\bibfield  {title} {\bibinfo {title}
  {Observational evidence from supernovae for an accelerating universe and a
  cosmological constant},\ }\href {https://doi.org/10.1086/300499} {\bibfield
  {journal} {\bibinfo  {journal} {Astron. J.}\ }\textbf {\bibinfo {volume}
  {116}},\ \bibinfo {pages} {1009} (\bibinfo {year} {1998})},\ \Eprint
  {https://arxiv.org/abs/astro-ph/9805201} {arXiv:astro-ph/9805201 [astro-ph]}
  \BibitemShut {NoStop}%
\bibitem [{\citenamefont {Perlmutter}\ \emph {et~al.}(1999)\citenamefont
  {Perlmutter} \emph {et~al.}}]{Perlmutter1999}%
  \BibitemOpen
  \bibfield  {author} {\bibinfo {author} {\bibfnamefont {S.}~\bibnamefont
  {Perlmutter}} \emph {et~al.},\ }\bibfield  {title} {\bibinfo {title}
  {Measurements of omega and lambda from 42 high-redshift supernovae},\ }\href
  {https://doi.org/10.1086/307221} {\bibfield  {journal} {\bibinfo  {journal}
  {Astrophys. J.}\ }\textbf {\bibinfo {volume} {517}},\ \bibinfo {pages} {565}
  (\bibinfo {year} {1999})},\ \Eprint {https://arxiv.org/abs/astro-ph/9812133}
  {arXiv:astro-ph/9812133 [astro-ph]} \BibitemShut {NoStop}%
\bibitem [{\citenamefont {{Planck Collaboration}}(2020)}]{2020}%
  \BibitemOpen
  \bibfield  {author} {\bibinfo {author} {\bibnamefont {{Planck
  Collaboration}}},\ }\bibfield  {title} {\bibinfo {title} {Planck 2018
  results. vi. cosmological parameters},\ }\href
  {https://doi.org/10.1051/0004-6361/201833910} {\bibfield  {journal} {\bibinfo
   {journal} {Astron. Astrophys.}\ }\textbf {\bibinfo {volume} {641}},\
  \bibinfo {pages} {A6} (\bibinfo {year} {2020})},\ \Eprint
  {https://arxiv.org/abs/1807.06209} {arXiv:1807.06209 [astro-ph.CO]}
  \BibitemShut {NoStop}%
\bibitem [{\citenamefont {Weinberg}(1989)}]{Weinberg1989}%
  \BibitemOpen
  \bibfield  {author} {\bibinfo {author} {\bibfnamefont {S.}~\bibnamefont
  {Weinberg}},\ }\bibfield  {title} {\bibinfo {title} {The cosmological
  constant problem},\ }\href {https://doi.org/10.1103/RevModPhys.61.1}
  {\bibfield  {journal} {\bibinfo  {journal} {Rev. Mod. Phys.}\ }\textbf
  {\bibinfo {volume} {61}},\ \bibinfo {pages} {1} (\bibinfo {year}
  {1989})}\BibitemShut {NoStop}%
\bibitem [{\citenamefont {{DESI
  Collaboration}}(2025{\natexlab{a}})}]{DESIDR2Cosmology}%
  \BibitemOpen
  \bibfield  {author} {\bibinfo {author} {\bibnamefont {{DESI
  Collaboration}}},\ }\bibfield  {title} {\bibinfo {title} {{DESI DR2 Results
  II: Measurements of Baryon Acoustic Oscillations and Cosmological
  Constraints}},\ }\href {https://doi.org/10.1103/tr6y-kpc6} {\bibfield
  {journal} {\bibinfo  {journal} {Phys. Rev. D}\ }\textbf {\bibinfo {volume}
  {112}},\ \bibinfo {pages} {083515} (\bibinfo {year} {2025}{\natexlab{a}})},\
  \Eprint {https://arxiv.org/abs/2503.14738} {arXiv:2503.14738 [astro-ph.CO]}
  \BibitemShut {NoStop}%
\bibitem [{\citenamefont {{DESI
  Collaboration}}(2025{\natexlab{b}})}]{DESIDR2ExtendedDE}%
  \BibitemOpen
  \bibfield  {author} {\bibinfo {author} {\bibnamefont {{DESI
  Collaboration}}},\ }\bibfield  {title} {\bibinfo {title} {Extended dark
  energy analysis using DESI DR2 BAO measurements},\ }\href
  {https://doi.org/10.1103/w4c6-1r5j} {\bibfield  {journal} {\bibinfo
  {journal} {Phys. Rev. D}\ }\textbf {\bibinfo {volume} {112}},\ \bibinfo
  {pages} {083511} (\bibinfo {year} {2025}{\natexlab{b}})},\ \Eprint
  {https://arxiv.org/abs/2503.14743} {arXiv:2503.14743 [astro-ph.CO]}
  \BibitemShut {NoStop}%
\bibitem [{\citenamefont {Gu}\ \emph {et~al.}(2025)\citenamefont {Gu},
  \citenamefont {Wang}, \citenamefont {Wang}, \citenamefont {Zhao} \emph
  {et~al.}}]{Gu_2025}%
  \BibitemOpen
  \bibfield  {author} {\bibinfo {author} {\bibfnamefont {G.}~\bibnamefont
  {Gu}}, \bibinfo {author} {\bibfnamefont {X.}~\bibnamefont {Wang}}, \bibinfo
  {author} {\bibfnamefont {Y.}~\bibnamefont {Wang}}, \bibinfo {author}
  {\bibfnamefont {G.-B.}\ \bibnamefont {Zhao}}, \emph {et~al.},\ }\bibfield
  {title} {\bibinfo {title} {Dynamical dark energy in light of the DESI DR2
  baryonic acoustic oscillations measurements},\ }\href
  {https://doi.org/10.1038/s41550-025-02669-6} {\bibfield  {journal} {\bibinfo
  {journal} {Nature Astronomy}\ }\textbf {\bibinfo {volume} {9}},\ \bibinfo
  {pages} {1879} (\bibinfo {year} {2025})},\ \Eprint
  {https://arxiv.org/abs/2504.06118} {arXiv:2504.06118 [astro-ph.CO]}
  \BibitemShut {NoStop}%
\bibitem [{\citenamefont {Raveri}\ \emph
  {et~al.}(2017)\citenamefont {Raveri}, \citenamefont {Bull},
  \citenamefont {Silvestri},\ and\ \citenamefont {Pogosian}}]{Raveri2017}%
  \BibitemOpen
  \bibfield  {author} {\bibinfo {author} {\bibfnamefont {M.}~\bibnamefont
  {Raveri}}, \bibinfo {author} {\bibfnamefont {P.}~\bibnamefont {Bull}},
  \bibinfo {author} {\bibfnamefont {A.}~\bibnamefont {Silvestri}},\ and\
  \bibinfo {author} {\bibfnamefont {L.}~\bibnamefont {Pogosian}},\ }\bibfield
  {title} {\bibinfo {title} {Priors on the effective dark energy equation of
  state in scalar-tensor theories},\ }\href
  {https://doi.org/10.1103/PhysRevD.96.083509} {\bibfield  {journal} {\bibinfo
  {journal} {Phys. Rev. D}\ }\textbf {\bibinfo {volume} {96}},\ \bibinfo
  {pages} {083509} (\bibinfo {year} {2017})},\ \Eprint
  {https://arxiv.org/abs/1703.05297} {arXiv:1703.05297 [astro-ph.CO]}


\bibitem{Phillips1993}%
M.~M.~Phillips,
``The Absolute Magnitudes of Type IA Supernovae,''
\href{https://doi.org/10.1086/186970}
{Astrophys. J. Lett. \textbf{413}, L105 (1993)}.

\bibitem{Guy2007}%
J.~Guy \emph{et al.},
``SALT2: Using distant supernovae to improve the use of Type Ia supernovae as distance indicators,''
\href{https://doi.org/10.1051/0004-6361:20066930}
{Astron. Astrophys. \textbf{466}, 11 (2007)},
\href{https://arxiv.org/abs/astro-ph/0701828}
{arXiv:astro-ph/0701828}.




\bibitem{Scolnic2018}%
D.~Scolnic \emph{et al.},
``The Complete Light-curve Sample of Spectroscopically Confirmed Type Ia Supernovae from Pan-STARRS1 and the Pantheon Sample,''
\href{https://doi.org/10.3847/1538-4357/aab9bb}
{Astrophys. J. \textbf{859}, 101 (2018)},
\href{https://arxiv.org/abs/1710.00845}
{arXiv:1710.00845}.

\bibitem{Brout2022}%
D.~Brout \emph{et al.},
``The Pantheon+ Analysis: Cosmological Constraints,''
\href{https://doi.org/10.3847/1538-4357/ac8e04}
{Astrophys. J. \textbf{938}, 110 (2022)},
\href{https://arxiv.org/abs/2202.04077}
{arXiv:2202.04077}.

\bibitem{Kelly2010}%
P.~L.~Kelly \emph{et al.},
``Hubble Residuals and Host Galaxy Properties in the Nearby Supernova Factory,''
\href{https://doi.org/10.1088/0004-637X/715/2/743}
{Astrophys. J. \textbf{715}, 743 (2010)},
\href{https://arxiv.org/abs/0912.0929}
{arXiv:0912.0929}.

\bibitem{Childress2014}%
M.~Childress \emph{et al.},
``Host Galaxy Properties and Hubble Residuals of Type Ia Supernovae,''
\href{https://doi.org/10.1088/0004-637X/770/2/108}
{Astrophys. J. \textbf{770}, 108 (2014)},
\href{https://arxiv.org/abs/1409.0009}
{arXiv:1409.0009}.
\bibitem [{\citenamefont {Schutz}(1986)}]{Schutz1986}%
  \BibitemOpen
  \bibfield  {author} {\bibinfo {author} {\bibfnamefont {B.~F.}\ \bibnamefont
  {Schutz}},\ }\bibfield  {title} {\bibinfo {title} {Determining the hubble
  constant from gravitational wave observations},\ }\href
  {https://doi.org/10.1038/323310a0} {\bibfield  {journal} {\bibinfo  {journal}
  {Nature}\ }\textbf {\bibinfo {volume} {323}},\ \bibinfo {pages} {310}
  (\bibinfo {year} {1986})}\BibitemShut {NoStop}%
\bibitem [{\citenamefont {Holz}\ and\ \citenamefont
  {Hughes}(2005)}]{HolzHughes2005}%
  \BibitemOpen
  \bibfield  {author} {\bibinfo {author} {\bibfnamefont {D.~E.}\ \bibnamefont
  {Holz}}\ and\ \bibinfo {author} {\bibfnamefont {S.~A.}\ \bibnamefont
  {Hughes}},\ }\bibfield  {title} {\bibinfo {title} {Using gravitational-wave
  standard sirens},\ }\href {https://doi.org/10.1086/431341} {\bibfield
  {journal} {\bibinfo  {journal} {Astrophys. J.}\ }\textbf {\bibinfo {volume}
  {629}},\ \bibinfo {pages} {15} (\bibinfo {year} {2005})},\ \Eprint
  {https://arxiv.org/abs/astro-ph/0504616} {arXiv:astro-ph/0504616 [astro-ph]}
  \BibitemShut {NoStop}%
\bibitem [{\citenamefont {{LIGO Scientific Collaboration}}\ \emph
  {et~al.}(2017)\citenamefont {{LIGO Scientific Collaboration}}, \citenamefont
  {{Virgo Collaboration}} \emph {et~al.}}]{Abbott2017GW170817}%
  \BibitemOpen
  \bibfield  {author} {\bibinfo {author} {\bibnamefont {{LIGO Scientific
  Collaboration}}}, \bibinfo {author} {\bibnamefont {{Virgo Collaboration}}},
  \emph {et~al.},\ }\bibfield  {title} {\bibinfo {title} {{GW170817}:
  Observation of gravitational waves from a binary neutron star inspiral},\ }
  \href {https://doi.org/10.1103/PhysRevLett.119.161101} {\bibfield  {journal}
  {\bibinfo  {journal} {Phys. Rev. Lett.}\ }\textbf {\bibinfo {volume} {119}},\
  \bibinfo {pages} {161101} (\bibinfo {year} {2017})},\ \Eprint
  {https://arxiv.org/abs/1710.05832} {arXiv:1710.05832 [gr-qc]}
  \BibitemShut {NoStop}%
\bibitem [{\citenamefont {{LIGO Scientific Collaboration}}\ \emph
  {et~al.}(2017)\citenamefont {{LIGO Scientific Collaboration}}, \citenamefont
  {{Virgo Collaboration}} \emph {et~al.}}]{Abbott2017StandardSiren}%
  \BibitemOpen
  \bibfield  {author} {\bibinfo {author} {\bibnamefont {{LIGO Scientific
  Collaboration}}}, \bibinfo {author} {\bibnamefont {{Virgo Collaboration}}},
  \emph {et~al.},\ }\bibfield  {title} {\bibinfo {title} {A gravitational-wave
  standard siren measurement of the Hubble constant},\ }\href
  {https://doi.org/10.1038/nature24471} {\bibfield  {journal} {\bibinfo
  {journal} {Nature}\ }\textbf {\bibinfo {volume} {551}},\ \bibinfo {pages}
  {85} (\bibinfo {year} {2017})},\ \Eprint {https://arxiv.org/abs/1710.05835}
  {arXiv:1710.05835 [astro-ph.CO]} \BibitemShut {NoStop}%
\bibitem [{\citenamefont {Zhao}\ \emph
  {et~al.}(2011)\citenamefont {Zhao}, \citenamefont {Van
  Den~Broeck}, \citenamefont {Baskaran},\ and\ \citenamefont {Li}}]{Zhao2011}%
  \BibitemOpen
  \bibfield  {author} {\bibinfo {author} {\bibfnamefont {W.}~\bibnamefont
  {Zhao}}, \bibinfo {author} {\bibfnamefont {C.}~\bibnamefont {Van
  Den~Broeck}}, \bibinfo {author} {\bibfnamefont {D.}~\bibnamefont
  {Baskaran}},\ and\ \bibinfo {author} {\bibfnamefont {T.~G.~F.}\ \bibnamefont
  {Li}},\ }\bibfield  {title} {\bibinfo {title} {Determination of dark energy
  by the Einstein Telescope: Comparing with CMB, BAO, and SNIa observations},\
  }\href {https://doi.org/10.1103/PhysRevD.83.023005} {\bibfield  {journal}
  {\bibinfo  {journal} {Phys. Rev. D}\ }\textbf {\bibinfo {volume} {83}},\
  \bibinfo {pages} {023005} (\bibinfo {year} {2011})},\ \Eprint
  {https://arxiv.org/abs/1009.0206} {arXiv:1009.0206 [astro-ph.CO]}
  \BibitemShut {NoStop}%
\bibitem [{\citenamefont {Maggiore}\ \emph
  {et~al.}(2020)\citenamefont {Maggiore} \emph
  {et~al.}}]{ETDesign}%
  \BibitemOpen
  \bibfield  {author} {\bibinfo {author} {\bibfnamefont {M.}~\bibnamefont
  {Maggiore}} \emph {et~al.},\ }\bibfield  {title} {\bibinfo {title} {Science
  case for the Einstein Telescope},\ }\href {https://doi.org/10.1088/1475-7516/2020/03/050}
  {\bibfield {journal} {\bibinfo {journal} {JCAP}\ }\bibinfo {volume} {03} (\bibinfo {year} {2020}) \bibinfo {pages} {050}},\
  \Eprint
  {https://arxiv.org/abs/1912.02622} {arXiv:1912.02622 [astro-ph.CO]}
  \BibitemShut {NoStop}%
\bibitem [{\citenamefont {Branchesi}\ \emph {et~al.}(2023)\citenamefont
  {Branchesi} \emph {et~al.}}]{Branchesi2023}%
  \BibitemOpen
  \bibfield  {author} {\bibinfo {author} {\bibfnamefont {M.}~\bibnamefont
  {Branchesi}} \emph {et~al.},\ }\bibfield  {title} {\bibinfo {title} {Science
  with the Einstein Telescope: A comparison of different designs},\ }\href {https://doi.org/10.1088/1475-7516/2023/07/068}
  {\bibfield {journal} {\bibinfo {journal} {JCAP}\ }\bibinfo {volume} {07} (\bibinfo {year} {2023}) \bibinfo {pages} {068}},\
  \Eprint
  {https://arxiv.org/abs/2303.15923} {arXiv:2303.15923 [gr-qc]} \BibitemShut
  {NoStop}%
\bibitem [{\citenamefont {Chevallier}\ and\ \citenamefont
  {Polarski}(2001)}]{Chevallier2001}%
  \BibitemOpen
  \bibfield  {author} {\bibinfo {author} {\bibfnamefont {M.}~\bibnamefont
  {Chevallier}}\ and\ \bibinfo {author} {\bibfnamefont {D.}~\bibnamefont
  {Polarski}},\ }\bibfield  {title} {\bibinfo {title} {Accelerating universes
  with scaling dark matter},\ }\href
  {https://doi.org/10.1142/S0218271801000822} {\bibfield  {journal} {\bibinfo
  {journal} {Int. J. Mod. Phys. D}\ }\textbf {\bibinfo {volume} {10}},\
  \bibinfo {pages} {213} (\bibinfo {year} {2001})},\ \Eprint
  {https://arxiv.org/abs/gr-qc/0009008} {arXiv:gr-qc/0009008 [gr-qc]}
  \BibitemShut {NoStop}%
\bibitem [{\citenamefont {Linder}(2003)}]{Linder2003}%
  \BibitemOpen
  \bibfield  {author} {\bibinfo {author} {\bibfnamefont {E.~V.}\ \bibnamefont
  {Linder}},\ }\bibfield  {title} {\bibinfo {title} {Exploring the expansion
  history of the universe},\ }\href
  {https://doi.org/10.1103/PhysRevLett.90.091301} {\bibfield  {journal}
  {\bibinfo  {journal} {Phys. Rev. Lett.}\ }\textbf {\bibinfo {volume} {90}},\
  \bibinfo {pages} {091301} (\bibinfo {year} {2003})},\ \Eprint
  {https://arxiv.org/abs/astro-ph/0208512} {arXiv:astro-ph/0208512 [astro-ph]}
  \BibitemShut {NoStop}%
\bibitem [{\citenamefont {Sathyaprakash}\ and\ \citenamefont
  {Schutz}(2009)}]{Sathyaprakash_2009}%
  \BibitemOpen
  \bibfield  {author} {\bibinfo {author} {\bibfnamefont {B.~S.}\ \bibnamefont
  {Sathyaprakash}}\ and\ \bibinfo {author} {\bibfnamefont {B.~F.}\ \bibnamefont
  {Schutz}},\ }\bibfield  {title} {\bibinfo {title} {Physics, astrophysics and
  cosmology with gravitational waves},\ }\href
  {https://doi.org/10.12942/lrr-2009-2} {\bibfield  {journal} {\bibinfo
  {journal} {Living Rev. Relativ.}\ }\textbf {\bibinfo {volume} {12}},\
  \bibinfo {pages} {2} (\bibinfo {year} {2009})},\ \Eprint
  {https://arxiv.org/abs/0903.0338} {arXiv:0903.0338 [gr-qc]} \BibitemShut
  {NoStop}%
\bibitem [{\citenamefont {Cai}\ and\ \citenamefont {Yang}(2017)}]{Cai_2017}%
  \BibitemOpen
  \bibfield  {author} {\bibinfo {author} {\bibfnamefont {R.-G.}\ \bibnamefont
  {Cai}}\ and\ \bibinfo {author} {\bibfnamefont {T.}~\bibnamefont {Yang}},\
  }\bibfield  {title} {\bibinfo {title} {Estimating cosmological parameters by
  the simulated data of gravitational waves from the Einstein Telescope},\
  }\href {https://doi.org/10.1103/PhysRevD.95.044024} {\bibfield  {journal}
  {\bibinfo  {journal} {Phys. Rev. D}\ }\textbf {\bibinfo {volume} {95}},\
  \bibinfo {pages} {044024} (\bibinfo {year} {2017})},\ \Eprint
  {https://arxiv.org/abs/1608.08008} {arXiv:1608.08008 [astro-ph.CO]}
  \BibitemShut {NoStop}%
\bibitem [{\citenamefont {Blanchet}(2014)}]{Blanchet2014}%
  \BibitemOpen
  \bibfield  {author} {\bibinfo {author} {\bibfnamefont {L.}~\bibnamefont
  {Blanchet}},\ }\bibfield  {title} {\bibinfo {title} {Gravitational
  radiation from post-Newtonian sources and inspiralling compact binaries},\ }
  \href {https://doi.org/10.12942/lrr-2014-2} {\bibfield  {journal} {\bibinfo
  {journal} {Living Rev. Relativ.}\ }\textbf {\bibinfo {volume} {17}},\
  \bibinfo {pages} {2} (\bibinfo {year} {2014})},\ \Eprint
  {https://arxiv.org/abs/1310.1528} {arXiv:1310.1528 [gr-qc]}
  \BibitemShut {NoStop}%
\bibitem [{\citenamefont {Schöneberg}(2026)}]{Schoneberg2026}%
  \BibitemOpen
  \bibfield  {author} {\bibinfo {author} {\bibfnamefont {N.}~\bibnamefont
  {Schöneberg}},\ }\href@noop {} {\bibinfo {title} {The 2024 {BBN} baryon
  abundance update}} (\bibinfo {year} {2026}),\ \Eprint
  {https://arxiv.org/abs/2401.15054} {arXiv:2401.15054 [astro-ph.CO]}
  \BibitemShut {NoStop}%
\bibitem [{\citenamefont {Rosenberg}\ \emph {et~al.}(2022)\citenamefont
  {Rosenberg}, \citenamefont {Gratton},\ and\ \citenamefont
  {Efstathiou}}]{Rosenberg_2022}%
  \BibitemOpen
  \bibfield  {author} {\bibinfo {author} {\bibfnamefont {E.}~\bibnamefont
  {Rosenberg}}, \bibinfo {author} {\bibfnamefont {S.}~\bibnamefont
  {Gratton}},\ and\ \bibinfo {author} {\bibfnamefont {G.}~\bibnamefont
  {Efstathiou}},\ }\bibfield  {title} {\bibinfo {title} {{CMB} power spectra
  and cosmological parameters from {Planck PR4} with {CamSpec}},\ }\href
  {https://doi.org/10.1093/mnras/stac2744} {\bibfield  {journal} {\bibinfo
  {journal} {Mon. Not. R. Astron. Soc.}\ }\textbf {\bibinfo {volume} {517}},\
  \bibinfo {pages} {4620} (\bibinfo {year} {2022})}\BibitemShut {NoStop}%
\bibitem [{\citenamefont {Cutler}\ and\ \citenamefont
  {Holz}(2009)}]{Cutler_2009}%
  \BibitemOpen
  \bibfield  {author} {\bibinfo {author} {\bibfnamefont {C.}~\bibnamefont
  {Cutler}}\ and\ \bibinfo {author} {\bibfnamefont {D.~E.}\ \bibnamefont
  {Holz}},\ }\bibfield  {title} {\bibinfo {title} {Ultrahigh precision
  cosmology from gravitational waves},\ }\href
  {https://doi.org/10.1103/PhysRevD.80.104009} {\bibfield  {journal} {\bibinfo
  {journal} {Phys. Rev. D}\ }\textbf {\bibinfo {volume} {80}},\ \bibinfo
  {pages} {104009} (\bibinfo {year} {2009})},\ \Eprint
  {https://arxiv.org/abs/0906.3752} {arXiv:0906.3752 [astro-ph.CO]}
  \BibitemShut {NoStop}%
\bibitem{Hild_2011}%
S.~Hild \emph{et al.},
``Sensitivity studies for third-generation gravitational wave observatories,''
\href{https://doi.org/10.1088/0264-9381/28/9/094013}
{Class. Quantum Grav. \textbf{28}, 094013 (2011)}.
\bibitem{Cai_2017}%
R.-G.~Cai and T.~Yang,
``Estimating cosmological parameters by the simulated
data of gravitational waves from the Einstein Telescope,''
\href{https://doi.org/10.1103/PhysRevD.95.044024}
{Phys. Rev. D \textbf{95}, 044024 (2017)}.
\bibitem [{\citenamefont {Ozel}\ and\ \citenamefont
  {Freire}(2016)}]{Ozel_2016}%
  \BibitemOpen
  \bibfield  {author} {\bibinfo {author} {\bibfnamefont {F.}~\bibnamefont
  {Ozel}}\ and\ \bibinfo {author} {\bibfnamefont {P.}~\bibnamefont {Freire}},\
  }\bibfield  {title} {\bibinfo {title} {Masses, radii, and the equation of
  state of neutron stars},\ }\href
  {https://doi.org/10.1146/annurev-astro-081915-023322} {\bibfield  {journal}
  {\bibinfo  {journal} {Annu. Rev. Astron. Astrophys.}\ }\textbf {\bibinfo
  {volume} {54}},\ \bibinfo {pages} {401} (\bibinfo {year} {2016})},\ \Eprint
  {https://arxiv.org/abs/1603.02698} {arXiv:1603.02698 [astro-ph.HE]}
  \BibitemShut {NoStop}%
\bibitem [{\citenamefont {Farrow}\ \emph {et~al.}(2019)\citenamefont {Farrow},
  \citenamefont {Zhu},\ and\ \citenamefont {Thrane}}]{Farrow_2019}%
  \BibitemOpen
  \bibfield  {author} {\bibinfo {author} {\bibfnamefont {N.}~\bibnamefont
  {Farrow}}, \bibinfo {author} {\bibfnamefont {X.-J.}\ \bibnamefont {Zhu}},\
  and\ \bibinfo {author} {\bibfnamefont {E.}~\bibnamefont {Thrane}},\
  }\bibfield  {title} {\bibinfo {title} {The mass distribution of galactic
  double neutron stars},\ }\href {https://doi.org/10.3847/1538-4357/ab12e3}
  {\bibfield  {journal} {\bibinfo  {journal} {Astrophys. J.}\ }\textbf
  {\bibinfo {volume} {876}},\ \bibinfo {pages} {18} (\bibinfo {year} {2019})},\
  \Eprint {https://arxiv.org/abs/1902.03300} {arXiv:1902.03300 [astro-ph.HE]}
  \BibitemShut {NoStop}%
\bibitem [{\citenamefont {Torrado}\ and\ \citenamefont
  {Lewis}(2021)}]{Torrado2021}%
  \BibitemOpen
  \bibfield {author} {\bibinfo {author} {\bibfnamefont {J.}~\bibnamefont
  {Torrado}}\ and\ \bibinfo {author} {\bibfnamefont {A.}~\bibnamefont
  {Lewis}},\ }\bibfield {title} {\bibinfo {title} {{Cobaya}: Code for
  Bayesian analysis of hierarchical physical models},\ }\href {https://doi.org/10.1088/1475-7516/2021/05/057}
  {\bibfield {journal} {\bibinfo {journal} {JCAP}\ }\bibinfo {volume} {05} (\bibinfo {year} {2021}) \bibinfo {pages} {057}},\
  \Eprint {https://arxiv.org/abs/2005.05290} {arXiv:2005.05290 [astro-ph.IM]}
  \BibitemShut {NoStop}%
\bibitem [{\citenamefont {Gelman}\ and\ \citenamefont
  {Rubin}(1992)}]{GelmanRubin}%
  \BibitemOpen
  \bibfield {author} {\bibinfo {author} {\bibfnamefont {A.}~\bibnamefont
  {Gelman}}\ and\ \bibinfo {author} {\bibfnamefont {D.~B.}\ \bibnamefont
  {Rubin}},\ }\bibfield {title} {\bibinfo {title} {Inference from iterative
  simulation using multiple sequences},\ }\href
  {https://doi.org/10.1214/ss/1177011136} {\bibfield {journal} {\bibinfo
  {journal} {Stat. Sci.}\ }\textbf {\bibinfo {volume} {7}},\ \bibinfo
  {pages} {457} (\bibinfo {year} {1992})}\BibitemShut {NoStop}%
\bibitem [{\citenamefont {Lewis}(2025)}]{Lewis_2019}%
  \BibitemOpen
  \bibfield  {author} {\bibinfo {author} {\bibfnamefont {A.}~\bibnamefont
  {Lewis}},\ }\bibfield  {title} {\bibinfo {title} {GetDist: A Python package
  for analysing Monte Carlo samples},\ }\href {https://doi.org/10.1088/1475-7516/2025/08/025}
  {\bibfield {journal} {\bibinfo {journal} {JCAP}\ }\bibinfo {volume} {08} (\bibinfo {year} {2025}) \bibinfo {pages} {025}},\
  \Eprint
  {https://arxiv.org/abs/1910.13970} {arXiv:1910.13970 [astro-ph.IM]}
  \BibitemShut {NoStop}%
\bibitem [{\citenamefont {Dembinski}\ \emph
  {et~al.}(2020)\citenamefont {Dembinski}, \citenamefont {Ongmongkolkul}
  \emph {et~al.}}]{Dembinski2020}%
  \BibitemOpen
  \bibfield {author} {\bibinfo {author} {\bibfnamefont {H.}~\bibnamefont
  {Dembinski}}, \bibinfo {author} {\bibfnamefont {P.}~\bibnamefont
  {Ongmongkolkul}}, \emph {et~al.},\ }\href
  {https://doi.org/10.5281/zenodo.3949207} {\bibinfo {title}
  {{scikit-hep/iminuit}}} (\bibinfo {year} {2020})\BibitemShut {NoStop}%
\bibitem [{\citenamefont {Wilks}(1938)}]{Wilks1938}%
  \BibitemOpen
  \bibfield  {author} {\bibinfo {author} {\bibfnamefont {S.~S.}\ \bibnamefont
  {Wilks}},\ }\bibfield  {title} {\bibinfo {title} {The large-sample
  distribution of the likelihood ratio for testing composite hypotheses},\
  }\href {https://doi.org/10.1214/aoms/1177732360} {\bibfield  {journal}
  {\bibinfo  {journal} {Ann. Math. Stat.}\ }\textbf {\bibinfo {volume} {9}},\
  \bibinfo {pages} {60} (\bibinfo {year} {1938})}\BibitemShut {NoStop}%
\bibitem [{\citenamefont {Cai}\ \emph {et~al.}(2010)\citenamefont {Cai},
  \citenamefont {Saridakis}, \citenamefont {Setare},\ and\ \citenamefont
  {Xia}}]{CaiQuintom2010}%
  \BibitemOpen
  \bibfield  {author} {\bibinfo {author} {\bibfnamefont {Y.-F.}\ \bibnamefont
  {Cai}}, \bibinfo {author} {\bibfnamefont {E.~N.}\ \bibnamefont {Saridakis}},
  \bibinfo {author} {\bibfnamefont {M.~R.}\ \bibnamefont {Setare}},\ and\
  \bibinfo {author} {\bibfnamefont {J.-Q.}\ \bibnamefont {Xia}},\ }\bibfield
  {title} {\bibinfo {title} {Quintom cosmology: Theoretical implications and
  observations},\ }\href {https://doi.org/10.1016/j.physrep.2010.04.001}
  {\bibfield  {journal} {\bibinfo  {journal} {Phys. Rep.}\ }\textbf {\bibinfo
  {volume} {493}},\ \bibinfo {pages} {1} (\bibinfo {year} {2010})},\ \Eprint
  {https://arxiv.org/abs/0909.2776} {arXiv:0909.2776 [hep-th]} \BibitemShut
  {NoStop}%
\bibitem [{\citenamefont {Chen}\ \emph {et~al.}(2024)\citenamefont {Chen},
  \citenamefont {Talbot},\ and\ \citenamefont {Chase}}]{ChenTalbotChase2024}%
  \BibitemOpen
  \bibfield  {author} {\bibinfo {author} {\bibfnamefont {H.-Y.}\ \bibnamefont
  {Chen}}, \bibinfo {author} {\bibfnamefont {C.}~\bibnamefont {Talbot}},\ and\
  \bibinfo {author} {\bibfnamefont {E.~A.}\ \bibnamefont {Chase}},\ }\bibfield
  {title} {\bibinfo {title} {Mitigating the counterpart selection effect for
  standard sirens},\ }\href {https://doi.org/10.1103/PhysRevLett.132.191003}
  {\bibfield  {journal} {\bibinfo  {journal} {Phys. Rev. Lett.}\ }\textbf
  {\bibinfo {volume} {132}},\ \bibinfo {pages} {191003} (\bibinfo {year}
  {2024})},\ \Eprint {https://arxiv.org/abs/2307.10402} {arXiv:2307.10402
  [astro-ph.CO]} \BibitemShut {NoStop}%
\bibitem [{\citenamefont {Holz}\ and\ \citenamefont
  {Linder}(2005)}]{HolzLinder2005}%
  \BibitemOpen
  \bibfield  {author} {\bibinfo {author} {\bibfnamefont {D.~E.}\ \bibnamefont
  {Holz}}\ and\ \bibinfo {author} {\bibfnamefont {E.~V.}\ \bibnamefont
  {Linder}},\ }\bibfield  {title} {\bibinfo {title} {Safety in numbers:
  Gravitational lensing degradation of the luminosity distance--redshift
  relation},\ }\href {https://doi.org/10.1086/432085} {\bibfield  {journal}
  {\bibinfo  {journal} {Astrophys. J.}\ }\textbf {\bibinfo {volume} {631}},\
  \bibinfo {pages} {678} (\bibinfo {year} {2005})},\ \Eprint
  {https://arxiv.org/abs/astro-ph/0412173} {arXiv:astro-ph/0412173}
  \BibitemShut {NoStop}%
\bibitem [{\citenamefont {Nicolaou}\ \emph {et~al.}(2020)\citenamefont
  {Nicolaou}, \citenamefont {Lahav}, \citenamefont {Lemos}, \citenamefont
  {Hartley},\ and\ \citenamefont {Braden}}]{Nicolaou2020}%
  \BibitemOpen
  \bibfield  {author} {\bibinfo {author} {\bibfnamefont {C.}~\bibnamefont
  {Nicolaou}}, \bibinfo {author} {\bibfnamefont {O.}~\bibnamefont {Lahav}},
  \bibinfo {author} {\bibfnamefont {P.}~\bibnamefont {Lemos}}, \bibinfo
  {author} {\bibfnamefont {W.}~\bibnamefont {Hartley}},\ and\ \bibinfo {author}
  {\bibfnamefont {J.}~\bibnamefont {Braden}},\ }\bibfield  {title} {\bibinfo
  {title} {The impact of peculiar velocities on the estimation of the hubble
  constant from gravitational wave standard sirens},\ }\href
  {https://doi.org/10.1093/mnras/staa1120} {\bibfield  {journal} {\bibinfo
  {journal} {Mon. Not. R. Astron. Soc.}\ }\textbf {\bibinfo {volume} {495}},\
  \bibinfo {pages} {90} (\bibinfo {year} {2020})},\ \Eprint
  {https://arxiv.org/abs/1909.09609} {arXiv:1909.09609 [astro-ph.CO]}
  \BibitemShut {NoStop}%
\bibitem [{\citenamefont {Essick}\ and\ \citenamefont
  {Holz}(2019)}]{EssickHolz2019}%
  \BibitemOpen
  \bibfield  {author} {\bibinfo {author} {\bibfnamefont {R.}~\bibnamefont
  {Essick}}\ and\ \bibinfo {author} {\bibfnamefont {D.~E.}\ \bibnamefont
  {Holz}},\ }\bibfield  {title} {\bibinfo {title} {Calibrating
  gravitational-wave detectors with {GW170817}},\ }\href
  {https://doi.org/10.1088/1361-6382/ab2142} {\bibfield  {journal} {\bibinfo
  {journal} {Class. Quantum Gravity}\ }\textbf {\bibinfo {volume} {36}},\
  \bibinfo {pages} {125002} (\bibinfo {year} {2019})},\ \Eprint
  {https://arxiv.org/abs/1902.08076} {arXiv:1902.08076 [astro-ph.IM]}
  \BibitemShut {NoStop}%
\bibitem [{\citenamefont {dos Santos}\ \emph {et~al.}(2025)}]{Santos2025}%
  \BibitemOpen
  \bibfield {author} {\bibinfo {author} {\bibfnamefont {F.~B.~M.}\ \bibnamefont {dos Santos}},
  \bibinfo {author} {\bibfnamefont {J.}~\bibnamefont {Morais}},
  \bibinfo {author} {\bibfnamefont {S.}~\bibnamefont {Pan}},
  \bibinfo {author} {\bibfnamefont {W.}~\bibnamefont {Yang}},\ and\
  \bibinfo {author} {\bibfnamefont {E.}~\bibnamefont {Di Valentino}},\ }
  \bibfield {title} {\bibinfo {title} {A New Window on Dynamical Dark Energy:
  Combining {DESI-DR2 BAO} with future Gravitational Wave Observations},\ }
  \href@noop {} {(\bibinfo {year} {2025})},\ \Eprint
  {https://arxiv.org/abs/2504.04646} {arXiv:2504.04646 [astro-ph.CO]}
  \BibitemShut {NoStop}%
\bibitem{Crittenden2012}%
R.~G.~Crittenden, G.-B.~Zhao, L.~Pogosian, L.~Samushia, and X.~Zhang,
``Fables of reconstruction: Controlling bias in the dark energy equation
of state,''
\href{https://doi.org/10.1088/1475-7516/2012/02/048}
{J. Cosmol. Astropart. Phys. \textbf{02} (2012) 048},
\Eprint{https://arxiv.org/abs/1112.1693}{arXiv:1112.1693 [astro-ph.CO]}.
\bibitem{Zhao_2017}%
G.-B.~Zhao \emph{et al.},
``Dynamical dark energy in light of the latest observations,''
\href{https://doi.org/10.1038/s41550-017-0216-z}
{Nat. Astron. \textbf{1}, 627 (2017)},
\Eprint{https://arxiv.org/abs/1701.08165}{arXiv:1701.08165 [astro-ph.CO]}.
\bibitem{Lewis_2000}%
A.~Lewis, A.~Challinor, and A.~Lasenby,
``Efficient computation of CMB anisotropies in closed FRW models,''
\href{https://doi.org/10.1086/309179}{Astrophys. J. \textbf{538}, 473
(2000)}.
\bibitem{Hahn_2023}%
C.~Hahn \emph{et al.},
``The DESI Bright Galaxy Survey: Final Target Selection, Design, and
Validation,''
\href{https://doi.org/10.3847/1538-3881/accff8}
{Astron. J. \textbf{165}, 253 (2023)}.
\bibitem{Zhou_2023}%
R.~Zhou \emph{et al.},
``Target Selection and Validation of DESI Luminous Red Galaxies,''
\href{https://doi.org/10.3847/1538-3881/aca5fb}
{Astron. J. \textbf{165}, 58 (2023)}.
\bibitem{Raichoor_2023}%
A.~Raichoor \emph{et al.},
``Target Selection and Validation of DESI Emission Line Galaxies,''
\href{https://doi.org/10.3847/1538-3881/acb213}
{Astron. J. \textbf{165}, 126 (2023)}.
\bibitem{Chaussidon_2023}%
E.~Chaussidon \emph{et al.},
``Target Selection and Validation of DESI Quasars,''
\href{https://doi.org/10.3847/1538-4357/acb3c2}
{Astrophys. J. \textbf{944}, 107 (2023)}.
\bibitem{Raveri_2019}%
M.~Raveri and W.~Hu,
``Concordance and discordance in cosmology,''
\href{https://doi.org/10.1103/PhysRevD.99.043506}
{Phys. Rev. D \textbf{99}, 043506 (2019)}.
\bibitem{Huterer_2003}%
D.~Huterer and G.~Starkman,
``Parametrization of dark-energy properties: A principal-component
approach,''
\href{https://doi.org/10.1103/PhysRevLett.90.031301}
{Phys. Rev. Lett. \textbf{90}, 031301 (2003)}.
\bibitem{Speagle2020}%
J.~S.~Speagle,
``{dynesty}: A dynamic nested sampling package for estimating Bayesian
posteriors and evidences,''
\href{https://doi.org/10.1093/mnras/staa278}
{Mon. Not. R. Astron. Soc. \textbf{493}, 3132 (2020)},
\Eprint{https://arxiv.org/abs/1904.02180}{arXiv:1904.02180 [astro-ph.IM]}.




\end{thebibliography}
\end{document}